\documentclass[%
 reprint,
superscriptaddress,
 amsmath,amssymb,
 aps,
]{revtex4-2}

\usepackage{graphicx}
\usepackage{dcolumn}
\usepackage{bm}

\graphicspath{{./figures/}}

\usepackage{color}
\definecolor{refcol}{RGB}{178,34,34}
\usepackage{microtype}
\usepackage[colorlinks,linkcolor=refcol,citecolor=refcol,urlcolor=refcol]{hyperref}

\definecolor{red}{rgb}{1,0,0}
\definecolor{blue}{rgb}{0,0,1}

\def\eq#1{(\ref{#1})}
\def\Eqs#1{Eqs.~(\ref{#1})}
\def\Eq#1{Eq.~(\ref{#1})}

\def\Fig#1{Fig.~\ref{#1}}
\def\Tab#1{Tab.~\ref{#1}}

\def\Sec#1{Sec.~\ref{#1}}

\begin{document}


\title{
Analytical structure of the equation of state at finite density:\\
Resummation versus expansion in a low energy model
}


\author{Swagato Mukherjee}
    \affiliation{Department of Physics, Brookhaven National Laboratory, Upton, New York 11973, USA}
    
    \author{Fabian Rennecke}
    \email{fabian.rennecke@theo.physik.uni-giessen.de} 
    \affiliation{Department of Physics, Brookhaven National Laboratory, Upton, New York 11973, USA}
    \affiliation{Institute for Theoretical Physics, Justus Liebig University Giessen, Heinrich-Buff-Ring 16, 35392 Giessen, Germany}
    \affiliation{Helmholtz Research Academy Hesse for FAIR (HFHF), Campus Giessen, 35392 Giessen, Germany}

    \author{Vladimir V. Skokov}
    \affiliation{Department of Physics, North Carolina State University, Raleigh, NC 27695, USA}
    \affiliation{RIKEN BNL Research Center, Brookhaven National Laboratory, Upton, New York 11973, USA}


\date{\today}
             
\begin{abstract}
For theories plagued with a sign problem at finite density, a Taylor expansion in the chemical potential is frequently used for lattice gauge theory based computations of the equation of state. Recently, in arXiv:2106.03165, a new resummation scheme was proposed for such an expansion that resums contributions of correlation functions of conserved currents to all orders in the chemical potential. Here, we study the efficacy of this resummation scheme using a low energy model, namely the mean-field quark-meson model. After adapting the scheme for a mean-field analysis, we confront the results of this scheme with the direct solution of the model at finite density as well as compare with results from Taylor expansions. We study to what extent the two methods capture the analytical properties of the equation of state in the complex chemical potential plane. As expected, the Taylor expansion breaks down as soon as the baryon chemical potential reaches the radius of convergence defined by the Yang-Lee edge singularity. Encouragingly, the resummation not only captures the location of the Yang-Lee edge singularity accurately, but is also able to describe the equation of state for larger chemical potentials beyond the location of the edge singularity for a wide range of temperatures.
\end{abstract}

\maketitle


\section{\label{sec:intro} Introduction}


Uncovering the structure of the phase diagram of Quantum Chromodynamics at nonzero temperature and density
has been the central goal for both the theoretical and the experimental nuclear physic community (see Ref.~\cite{An:2021wof} for a review). Non-perturbative theoretical understanding
of the QCD phase diagram is hampered by the so-called sign problem. To introduce it, we  consider a theory containing bosonic fields $\Phi$ and fermionic fields $\psi$ in Euclidean spacetime with the following classical action
\begin{align}\label{eq:Sgen}
S[\Phi,\psi,\bar\psi] = S_\Phi[\Phi] + \int_0^\beta\! dx_0 \int\! d^3x\, \bar \psi(x)\, M(\Phi;\mu)\, \psi(x)\,.
\end{align}
Here $S_\Phi$ is the part of the action that only depends on $\Phi$. $M(\Phi;\mu)$ is a Dirac operator which includes a coupling between  the fermionic and bosonic fields. We are considering finite temperature $T = 1/\beta $ and finite chemical potential $\mu$.  Thus the Dirac operator, as indicated above,  explicitly depends on the chemical potential $\mu$. A prominent example of such a theory is QCD with the bosonic fields to be identified with gluons and the fermionic fields with quarks.

The partition function of this theory can formally be obtained from the Euclidean path integral,
\begin{align}
Z = \int\!\mathcal{D}\Phi\,\mathcal{D}\psi\,\mathcal{D}\bar\psi\, e^{-S}\,.
\end{align}
The grand canonical thermodynamic potential $\Omega$ is then proportional to  $\ln Z$.
The action in \Eq{eq:Sgen} is quadratic in the fermionic fields; therefore they can be readily integrated out, resulting in the (non-local) functional fermionic determinant $\det M$, 
\begin{align}
Z = \int\!\mathcal{D}\Phi\, \exp\Big[-S_\Phi[\Phi] + \ln\det M(\Phi;\mu)\Big]\,.
\end{align}
In numerous theories, including QCD, the presence of a finite chemical potential in the Dirac operator gives rise to a sign problem: a real $\mu$ can lead to a complex spectrum of the Dirac operator. In this case the weight of the configurations of $\Phi$, $\sim \det M$, is complex, rendering the Monte-Carlo importance sampling impractical.

Aside from attempts to entirely circumvent the sign problem, e.g.\ by using methods that do not have to rely on importance sampling~\cite{Aarts:2009yj,
Aarts:2013uxa, Sexty:2013ica, Fodor:2015doa, Cristoforetti:2012su, Fukuma:2019uot, Alexandru:2020wrj, Fischer:2018sdj, Fu:2019hdw},  a common strategy is to expand the path integral about $\mu=0$~\cite{Bazavov:2017dus, Datta:2016ukp}. This yields a power series in $\mu$, where each coefficient can be computed from the path integral with the $\mu=0$ weight. Information at finite $\mu$ is therefore obtained through an extrapolation from $\mu=0$. 

{
Although improvements of the the conventional Taylor expansion have been proposed in the literature \cite{Borsanyi:2021sxv}, a major obstacle for schemes based on analytical expansions is that they are bound by the analytical constraints of the underlying theory. Singularities in the complex plane determine the radius of convergence of analytical expansions. Resummations based on Pad\'e approximations can be used to estimate the location of the (nearest) singularities, see, e.g., \cite{Gavai:2008zr, Karsch:2010hm, Datta:2016ukp, Basar:2021hdf, Schmidt:2021pey}. Still, precise knowledge of expansion coefficients of very high order is required and information beyond the singularities are difficult to obtain.

In this work, we study the approach introduced in \cite{Mondal:2021jxk}, where contributions of $(n\!\leq\! N)$-point correlation functions of the fermion number currents to the thermodynamic potential are resummed to all orders in $\mu$. In addition to improved convergence, zeros of the partition function of QCD at imaginary $\mu$ have been identified, which could be related to physical singularities of the thermodynamic potential. The main motivation of this work is to examine this in detail regarding the analytic and thermodynamic properties of a model where these quantities can be computed directly.

To this end, we use a quark-meson model which can be solved directly in mean-field approximation. While simplistic, this model captures some basic features of QCD at low energies and of the chiral phase transition, see, e.g., \cite{Jungnickel:1995fp, Schaefer:2004en, Skokov:2010sf, Pawlowski:2014zaa, Resch:2017vjs} for studies in mean-field and beyond. After introducing the resummation scheme and adapting it to a mean-field approximation in \Sec{sec:main}, we introduce the model in \Sec{sec:model}. In \Sec{sec:ana} we discuss the analytical properties of the model for complex chemical potential and study how well it can be reproduced by the resummation scheme. Full results for the thermodynamics at real and imaginary chemical potentials are confronted with the resummation and the conventional Taylor expansion in \Sec{sec:thermo}.
}

\section{\label{sec:main} Resummation}

We start by introducing the resummation scheme of Ref.\ \cite{Mondal:2021jxk}. To this end, we define the derivatives of the logarithm of the fermion determinant as
\begin{align}
D_n(\Phi) = \frac{\partial^n \ln\det M(\Phi;\mu)}{\partial\hat\mu^n}\bigg|_{\mu=0}\,,
\end{align}
with $\hat\mu = \mu/T$. Expanding the fermion determinant inside the path integral to order $N$ in powers of $\hat\mu$ yields
\begin{align}\label{eq:ZR}
\begin{split}
Z^R_N &= \int\!\mathcal{D}\Phi\,\exp\!\bigg[ \sum_{n=1}^N \frac{1}{n!} D_n(\Phi)\hat\mu^n \bigg] \det M(\Phi;0)\,e^{-S_\Phi[\Phi]}\\
&= \bigg\langle\!\exp\!\bigg[  \sum_{n=1}^N \frac{1}{n!} D_n(\Phi)\hat\mu^n\bigg] \bigg\rangle_{\!0}\,.
\end{split}
\end{align}
The ensemble average at $\mu=0$ is given by
\begin{align}
\langle A \rangle_0 = \int\!\mathcal{D}\Phi\, A\, \det M(\Phi;0)\, e^{-S_\Phi[\Phi]}.
\end{align}
The resulting thermodynamic potential is
\begin{align}\label{eq:OmegaR}
\Omega^R_N(T,\mu) = -\frac{T}{V} \ln Z^R_N\,,
\end{align}
where $V$ is the spatial volume.
Crucially, even at finite order of the expansion, $N$, $\Omega_N^R$ contains infinite powers of $\hat\mu$.

In contrast, an ordinary Taylor expansion of the thermodynamic potential about $\mu=0$ to order $N$,
\begin{align}\label{eq:OmegaE}
\Omega_N^E(T,\mu) = \sum_{n=1}^N \frac{1}{n!} \frac{\partial^n \Omega(T,\mu)}{\partial\hat\mu^n}\bigg|_{\hat\mu=0}\, \hat\mu^n\,,
\end{align}
is only an $N$-th order polynomial of $\hat\mu$ by construction. In theories with charge conjugation symmetry (including QCD), only even powers of $\hat\mu$ contribute in \Eq{eq:OmegaE}.
$\Omega_N^E$ can be expressed in terms of averages of the $D_n$ at $\mu=0$ \cite{Allton:2005gk}. As pointed out in \cite{Mondal:2021jxk}, $\Omega_N^R$ can be interpreted as an all-order resummation of finite-order contributions to $\Omega_N^E$.
This resummation is directly connected to the reweighting method; for recent developments see, e.g., \cite{Giordano:2020roi, Borsanyi:2021hbk}. Expanding the logarithm of the fermion determinant in the weight $\det M(\Phi;\mu)/\det M(\Phi;0)$ in powers of $\mu$ leads to \Eq{eq:ZR}.

Naturally, the nontrivial analytic structure of the thermodynamic potential in plane of complex $\mu$, see \cite{Yang:1952be, Lee:1952ig}, cannot be captured by a strictly analytic expansion in $\mu$. Yet, the closest singularity in the complex plane determines the radius of convergence of the expansion. Both $\Omega_N^R$ and $\Omega_N^E$ can be evaluated at complex $\mu$. However, unlike $\Omega_N^E$, which cannot resolve such singularities (directly), ${\rm Re}[Z^R_N]$ in \Eq{eq:ZR} can become negative, resulting in a singular $\Omega_N^R$.

In this work, we want to test \Eq{eq:OmegaR} using a  mean-field approximation, i.e.\ to leading order in the saddle point approximation of the path integral. 
In general, the thermodynamic potential on a background field $\bar\Phi$ is
\begin{align}\label{eq:omegabar}
\bar\Omega(T,\mu;\bar\Phi)= -\frac{T}{V} \Big\{ S_\Phi[\bar\Phi]+\ln\det M(\bar\Phi;\mu)\Big\}\,,
\end{align}
and the stationary point $\bar\Phi_0$ is determined by  
\begin{align}
\frac{\delta \bar\Omega(T,\mu;\bar\Phi)}{\delta \bar\Phi} \bigg|_{\bar\Phi_0} = 0\,.
\end{align}
Correspondingly, \Eq{eq:OmegaR} becomes
\begin{align}\label{eq:barOmegaR}
\begin{split}
\bar\Omega_N^{R}(T,\mu;\bar\Phi)= -\frac{T}{V} \bigg\{& S_\Phi[\bar\Phi]+\ln\det M(\bar\Phi;0)\\
&+\sum_{n=1}^N \frac{1}{n!} D_n(\bar\Phi)\hat\mu^n\bigg\}\,.
\end{split}
\end{align}
Since the expansion in $\hat\mu$ in \Eq{eq:ZR} is done before the ensemble average, the stationary point $\bar\Phi_0^R$ is defined by \Eq{eq:barOmegaR},
\begin{align}\label{eq:REoM}
\frac{\delta \bar\Omega_N^{R}(T,\mu;\bar\Phi)}{\delta \bar\Phi} \bigg|_{\bar\Phi_0^R} = 0\,,
\end{align}
and all thermodynamic quantities can be extracted from
\begin{align}
\bar\Omega_N^{R}(T,\mu) \equiv \bar\Omega_N^{R}(T,\mu;\bar\Phi_0^R)\,.
\end{align}
Owing to the explicit $\mu$-dependence of $\bar\Omega_N^{R}(T,\mu;\bar\Phi)$, the stationary point of the resummed expansion depends nontrivially on $\mu$, $\bar\Phi_0^R = \bar\Phi_0^R(\mu)$. Through this dependence, $\bar\Omega_{N}^{R}$ is in general a nonanalytic function of $\mu$.

This is in contrast to the ordinary Taylor expansion in mean-field,
\begin{align}\label{eq:barOmegaE}
\bar\Omega_N^{E}(T,\mu) = -\sum_{n=0}^N \frac{\chi_n(0)}{n!}\, \hat\mu^n\,,
\end{align}
where the coefficients are given by the the susceptibilities,
\begin{align}\label{eq:sus}
\chi_n(\mu) = \frac{T}{V} \frac{\partial^n \bar\Omega(T,\mu;\bar\Phi_0)}{\partial\hat\mu^n}\,,
\end{align}
evaluated at $\mu = 0$ and therefore do not depend on $\mu$. For any finite $N$, $\bar\Omega_N^{E}$ is a finite ($N$-th) order polynomial in $\mu^2$ and hence  $\bar\Omega_N^{E}$  is strictly analytic.

\section{\label{sec:model} Model}

To test the resummation scheme of \cite{Mondal:2021jxk} directly, we use a quark-meson model with $N_f=2$ degenerate quark flavors and  $N_c = 3$ colors. The Euclidean action is
\begin{align}\label{eq:lag}
\begin{split}
S^{\rm QM} &= \int_0^\beta\! dx_0 \int\! d^3x\,\bigg\{ \bar\psi \bigg( \gamma_\mu\partial_\mu+ \frac{1}{2} h\, \bm{\tau} \bm{\phi} + \gamma_0 \mu \bigg)\psi\\
&\quad + \frac{1}{2} (\partial_\mu \bm\phi)^2 + U(\bm{\phi}^2) - j \sigma\bigg\}\,.
\end{split}
\end{align}
$\gamma_\mu$ are the Euclidean gamma matrices, $\bm{\tau}^T = (1,i\gamma_5 \vec{\tau})$ with the Pauli matrices $\vec{\tau}$, and $\bm{\phi}^T = (\sigma,\vec{\pi})$ is the $O(4)$ meson field. $U(\bm{\phi}^2)$ is the $O(4)$ symmetric effective meson potential. An explicit symmetry breaking is introduced through the source $j$, which can be related  to the current quark mass. The precise form of this relation is of no importance for our study.

In this work we employ a mean-field approximation to compute the thermodynamic potential $\Omega$ based on \Eq{eq:lag}. Assuming a homogeneous mean field, the meson background field is
\begin{align}
\bar{\bm{\phi}}=
\begin{pmatrix}\bar\sigma \\ \vec{0}\end{pmatrix}\,,
\end{align}
resulting in the Dirac operator
\begin{align}\label{eq:DOQM}
M^{\rm QM}(\bar\sigma;\mu) = \gamma_\mu\partial_\mu + \gamma_0 \mu + \frac{1}{2} h \bar\sigma\,,
\end{align}
where the appropriate unit matrices in spinor-, color- and flavor space are implied. The thermodynamic potential then is
\begin{align}
\begin{split}
\bar\Omega^{\rm QM}(T,\mu;\bar\sigma) 
=U(\bar{\bm{\phi}}^2) - j\bar\sigma - \frac{T}{V} \ln\det M^{\rm QM}(\bar\sigma;\mu)\,,
\end{split}
\end{align}
The quark determinant $\det M$ can be evaluated using conventional methods of thermal field theory, see, e.g., \cite{Laine:2016hma}. With the fermionic Matsubara frequency $\nu_n = (2n+1) \pi T$ and the quark energy $E_q(\bar\sigma) = \sqrt{q^2+\frac{1}{4} h^2 \bar\sigma^2}$, where $q^2 \equiv \vec{q}^{\,2}$, it reads
\begin{align}
\begin{split}
 &\frac{T}{V} \ln\det M^{\rm QM}(\bar\sigma;\mu)\\
 &\;= 2 N_f N_c\, T\!\sum_{n=-\infty}^\infty\int_q \ln \Big[ (\nu_n + i\mu )^2 + E_q(\bar\sigma)^2 \Big]\\
 &\;\equiv 2 N_f N_c \Big[ J_0(\bar\sigma) + J_{T,\mu}(\bar\sigma) + J_{T,-\mu}(\bar\sigma)  \Big]\,,
\end{split}
\end{align}
where we used the shorthand notation $\int_q = \int\!\frac{d^3q}{(2\pi)^3}$. The thermal contribution to the quark determinant is given by
\begin{align}\label{eq:JT}
J_{T,\mu}(\bar\sigma) = \frac{1}{2 \pi^2} \int_0^\infty\!dq\, q^2\, T\, \ln\Big[ 1 + e^{-(E_q(\bar\sigma)-\mu)/T} \Big]\,.
\end{align}
Away from the low- and high-temperature limits, this integral has to be carried out numerically. The vacuum contribution $J_0$ is ultraviolet-divergent. Nonetheless, the finite piece of the vacuum contribution may and does depend on the meson field. This contribution has to carefully extracted~\cite{Skokov:2010sf}. For this we perform the expansion around $d=3-2\epsilon$ dimension. It yields
\begin{align}
J_0(\bar\sigma) = -\frac{h^4\bar\sigma^4 }{2^9 \pi^2 \Lambda^{2\epsilon}}
\bigg[ \frac{1}{\epsilon} - \ln\bigg(\frac{h^2\bar\sigma^2}{4 \Lambda^2}\bigg) +C+ \mathcal{O}(\epsilon)\bigg]\,,
\end{align}
with the constant $C = \ln(4\pi)-\gamma_E+\frac{3}{2}$, where $\gamma_E$ is the Euler Mascheroni constant. $\Lambda$ is the renormalization scale parameter. We subtract the divergent piece and the constant, take the limit $\epsilon\rightarrow 0$, and arrive at the vacuum contribution in dimensional regularization,
\begin{align}\label{eq:J0eps}
J_0^\epsilon(\bar\sigma) = \frac{h^4\bar\sigma^4}{2^9 \pi^2} \ln\bigg(\frac{h^2\bar\sigma^2}{4 \Lambda^2}\bigg)\,.
\end{align}
%
\begin{table}[t]
\renewcommand{\arraystretch}{1.3}
\begin{center}
\begin{tabular}{ c | l }
\hline \hline
$h$ & $2\times 300/93$ \\
$\lambda$ & 21 \\
$\nu$ & $106\;{\rm MeV}$\\
$j$ & $138^2\times 93\;{\rm MeV}^3$\\
$\Lambda$ & $500\;{\rm MeV}$ \\
\hline
$f_\pi$ & $92.67\;{\rm MeV}$ \\
$m_\pi$ & $138.25\;{\rm MeV}$ \\
$m_\sigma$ & $496.59\;{\rm MeV}$ \\
$T_{\rm pc}$ & $163.62\;{\rm MeV}$ \\
$(T_{\rm CEP},\,\mu_{B,{\rm CEP}})$ & $(27.46,\, 895.85)\;{\rm MeV}$ \\
\hline \hline
\end{tabular}
\end{center}
\caption{Model parameters (upper half) and resulting physical quantities (lower half). $T_{\rm pc}$ is the pseudocritical temperature at $\mu_B$ = 0, defined via the maximum of the correlation length $m_\sigma^{-1}$. $(T_{\rm CEP},\,\mu_{B,{\rm CEP}})$ is the location of the critical endpoint. $\mu_B = 3\mu$ is the baryon chemical potential.}
\label{tab:para}
\end{table}
%
For the symmetric meson potential we use an ansatz which allows for spontaneous symmetry breaking, 
\begin{align}
U(\bm{\phi}^2) = \frac{\lambda}{4} \big(\bm{\phi}^2-\nu^2\big)^2\,,
\end{align}
so that the regularized thermodynamic potential in the mean-field approximation becomes
\begin{align}\label{eq:barOmegaQM}
\begin{split}
\bar\Omega^{\rm QM}(T,\mu;\bar\sigma) &= \frac{\lambda}{4} (\bar\sigma^2-\nu^2)^2-j\bar\sigma\\
&\quad- 2 N_f N_c \Big[ J_0^\epsilon(\bar\sigma) + J_{T,\mu}(\bar\sigma) + J_{T,-\mu}(\bar\sigma) \Big]\,.
\end{split}
\end{align}
The vacuum contribution of the quarks is given in \Eq{eq:J0eps} and the thermal contribution in \Eq{eq:JT}.
Physical results are extracted the minimum of the thermodynamic potential, that is 
\begin{align}\label{eq:barOmegaQM2}
\bar\Omega^{\rm QM}(T,\mu)=\bar\Omega^{\rm QM}(T,\mu;\bar\sigma_0)\,,
\end{align}
where $\bar\sigma_0$ is the solution of the equation of motion,
\begin{align}\label{eq:QMEoM}
\frac{\partial \bar\Omega^{\rm QM}(T,\mu;\bar\sigma)}{\partial \bar\sigma}\bigg|_{\bar\sigma_0} = 0\,.
\end{align}
The influence of the vacuum contribution on the thermodynamics of the quark-meson model has been studied in \cite{Skokov:2010sf}.

We thus have all ingredients of the model fully determined, as \Eqs{eq:barOmegaQM} and \eq{eq:QMEoM} can numerically be solved for an arbitrary (complex) $\mu$. {
Using \Eqs{eq:barOmegaR} and \eq{eq:REoM} we compute the resummed thermodynamic potential $\bar\Omega_N^{\rm{QM},R}$. The corresponding expansion coefficients $D_n$ in \Eqs{eq:barOmegaR} are given by
\begin{align}
    D_n^{\rm QM}(\bar\sigma) = \frac{2 N_f N_c V}{T}\, \frac{\partial^n}{\partial \hat\mu^n} \big[ J_{T,\mu}(\bar\sigma) + J_{T,-\mu}(\bar\sigma)  \big]\Big|_{\mu = 0}\,,
\end{align}
where $J_{T,\mu}$ is defined in \Eq{eq:JT}.
Furthermore, using \Eqs{eq:barOmegaE} and \eq{eq:sus} we compute the expanded and truncated thermodynamic potential $\bar\Omega_N^{\rm{QM},E}$. 
This is done by first computing $\bar\Omega^{\rm QM}(T,\mu)$ in \Eq{eq:barOmegaQM2} for a small region of chemical potentials around $\mu=0$ and then taking numerical derivatives at $\mu=0$ to get the susceptibilities $\chi_n^{\rm QM}(0)$ defined in \Eq{eq:sus}.
}
Table \ref{tab:para} shows the model parameters and resulting physical quantities we use for the numerical analysis. We use the baryon chemical potential $\mu_B = 3 \mu$ in the following.

\section{\label{sec:ana} Analytic Structure}

%
\begin{figure}[t]
\centering
\includegraphics[width=.45\textwidth]{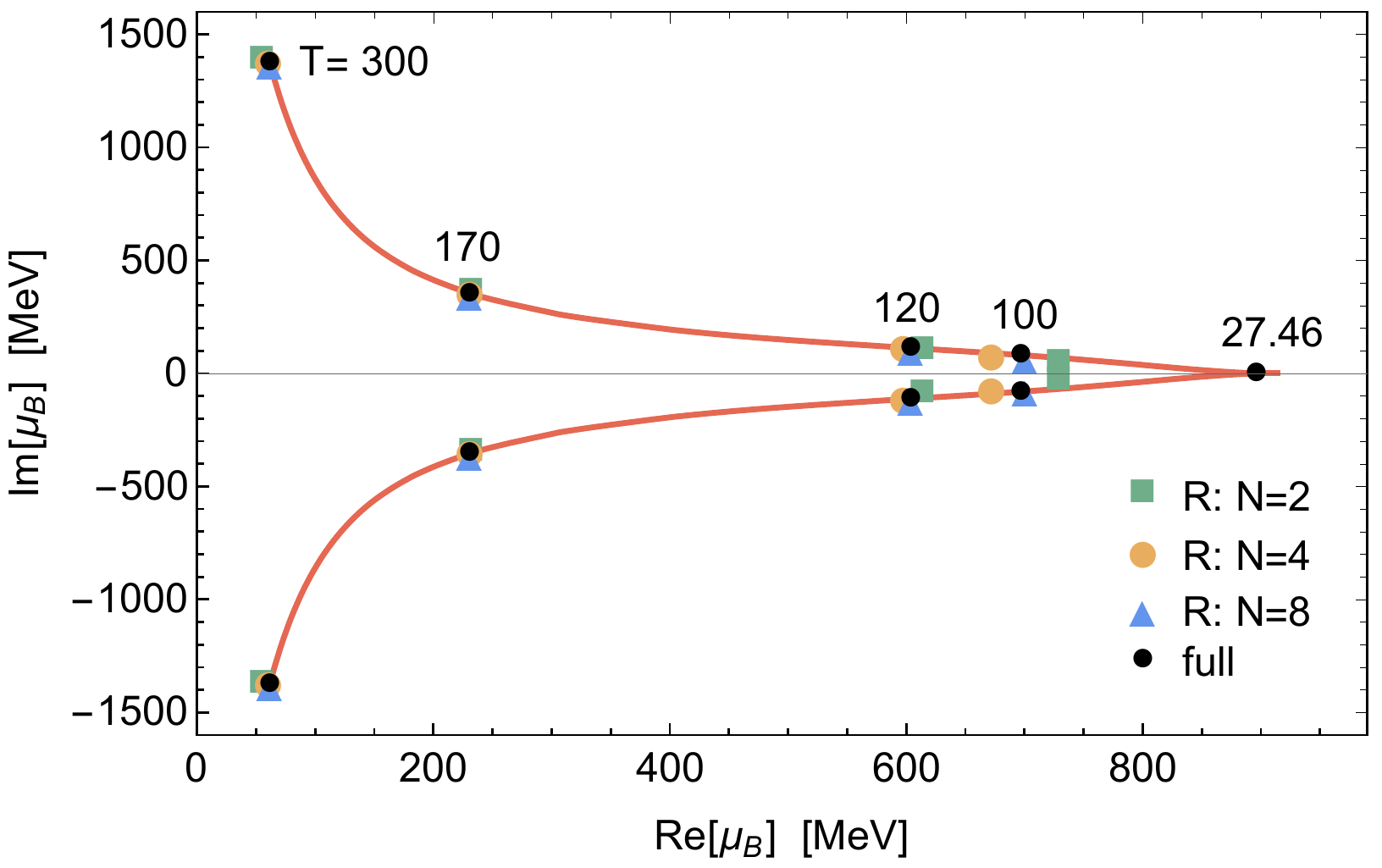}
\caption{Location of the Yang-Lee edge singularities in the complex right half plane of $\mu_B$ for $T = 1 - 300\,{\rm MeV}$. The red line shows the location of the full result.  The points illustrate the location for specific temperatures, where the full result is compared to the resummed result at different truncation orders.
For $T\lesssim 90$~MeV, the resummation method does not provide reliable results on the location of the edge singularity.
For $T < T_{\rm CEP} = 27.46\, {\rm MeV}$, instead of the edge singularity, there is a cut across the real axis, reflecting a first order phase transition.}
\label{fig:edge}
\end{figure}
%

We first study how well the different methods capture the analytical structure of the model 
in the complex $\mu_B$-plane. 
Phase transitions in the system are related to a branch cut in the thermodynamic potential at complex $\mu_B$. At a second (first) order phase transition this cut pinches (crosses) the real $\mu_B$ axis. In the symmetric phase above the phase transition, the cut terminates at complex conjugate branch points, known as the Yang-Lee edge singularity $\mu_B^{\rm YL}$ \cite{Yang:1952be, Lee:1952ig}. 
In the vicinity of the pseudocritical temperature the closest singularity to the origin $\mu_B=0$ is related to the second order phase transition in the chiral limit. Indeed, it has been shown by Fisher \cite{Fisher:1978pf} that, at a given temperature,  the Yang-Lee edge singularity corresponds to a critical point in the complex plane. For a more detailed discussion on this topic we refer to~\cite{Stephanov:2006dn}, see also \cite{Almasi:2019bvl,Mukherjee:2019eou,Connelly:2020gwa}.  
To find the edge singularity $\mu_B^{\rm YL}$, we can follow the same  procedure as when finding a critical point, that is  we solve
the system of equations 
\begin{align}\label{eq:findYLE}
\begin{split}
\frac{\partial \bar\Omega^{\rm QM}(T,\mu_B/3;\bar\sigma)}{\partial\bar\sigma} &= 0\,,\\[1ex]
\frac{\partial^2 \bar\Omega^{\rm QM}(T,\mu_B/3;\bar\sigma)}{\partial\bar\sigma^2} &= 0\,.
\end{split}
\end{align}
In contrast to a conventional critical point, where we look at real parameters $T$ and $\mu_B$, here, for a given $T$ we solve 
for real and imaginary parts of $\mu_B^{\rm YL}$ and  $\bar\sigma^{YL}$. Note that the system consists of four equations, as for complex $\mu_B$, the thermodynamic potential is complex-valued. The results for the location of the edge singularity  for temperatures between 1 and 300~MeV are shown in \Fig{fig:edge}.

The truncated Taylor expansion is analytic for any $N$ and thus cannot provide direct information about the location of the Yang-Lee edge (except through the analysis of the radius of the convergence of the expansion). In other words, for any given order $N$, there are no solutions of \Eq{eq:findYLE} with $\bar\Omega^{\rm QM} \to \bar\Omega^{\rm QM,E}_N$. 
In contrast to this, we find that the resummed thermodynamic potential can provide direct information about the location of the Yang-Lee edge singularity. In order to test this, we also solve \Eq{eq:findYLE} using $\bar\Omega^{\rm QM,R}_N$. Even at order $N=2$, the exact location is reproduced rather accurately. For a comparison of the result for different orders $N$ and the exact location at different temperatures, see \Fig{fig:edge}. In general, even the lowest order resummation gives precise results on the location of the edge singularity for $T \gtrsim 120$~MeV, and higher orders increase the precision of the result. The resummation converges a bit slower at lower temperatures, but accurate results can be achieved, e.g., for order $N=8$ at $T=100$~MeV. However, the resummation completely fails to describe the location of the edge singularity below $T \lesssim 90$~MeV. This is directly related to the absence of thermal cuts in the resummation scheme.

%
\begin{figure*}[t]
\centering
\includegraphics[width=.48\textwidth]{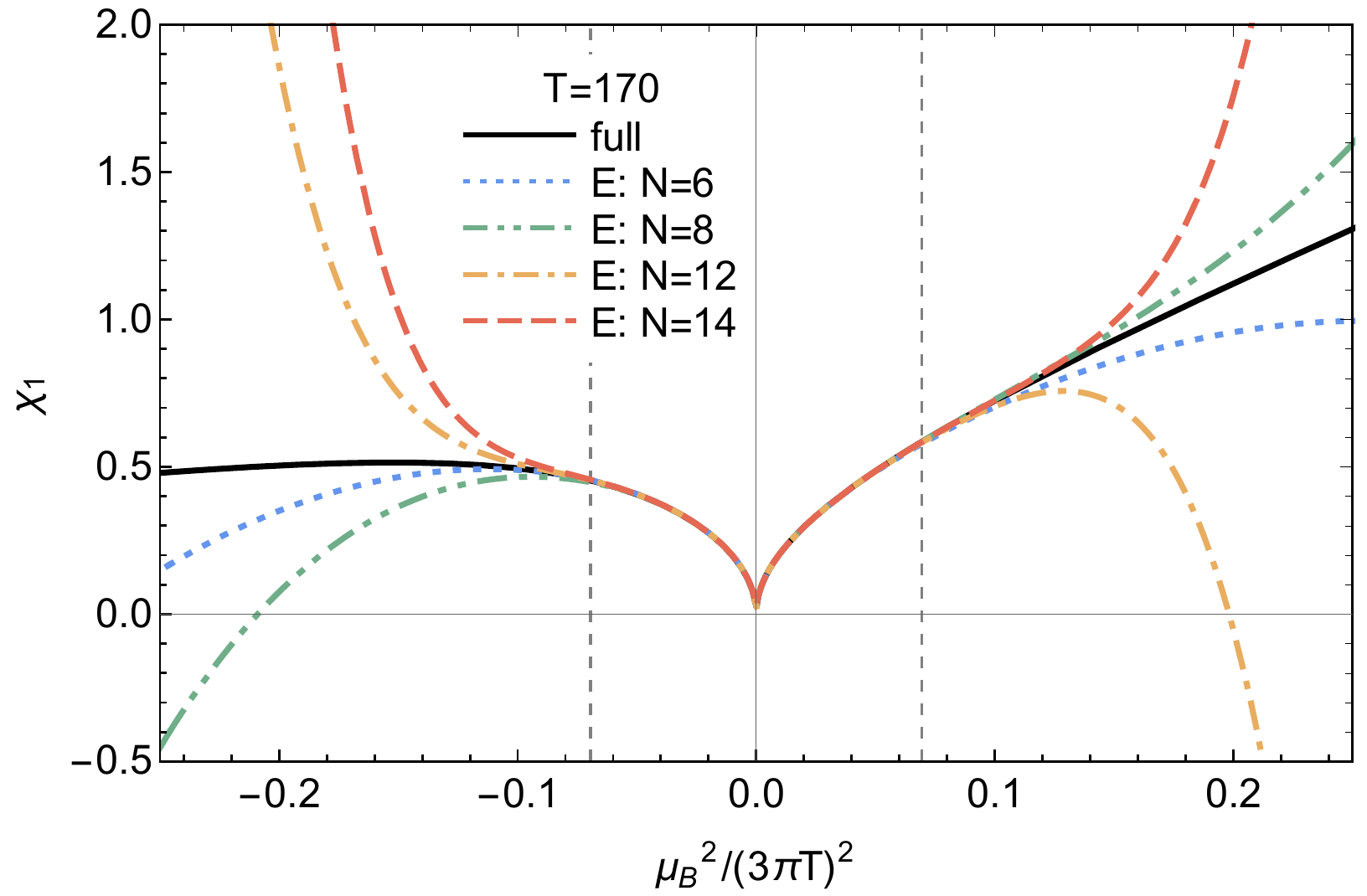}
\hfill
\includegraphics[width=.48\textwidth]{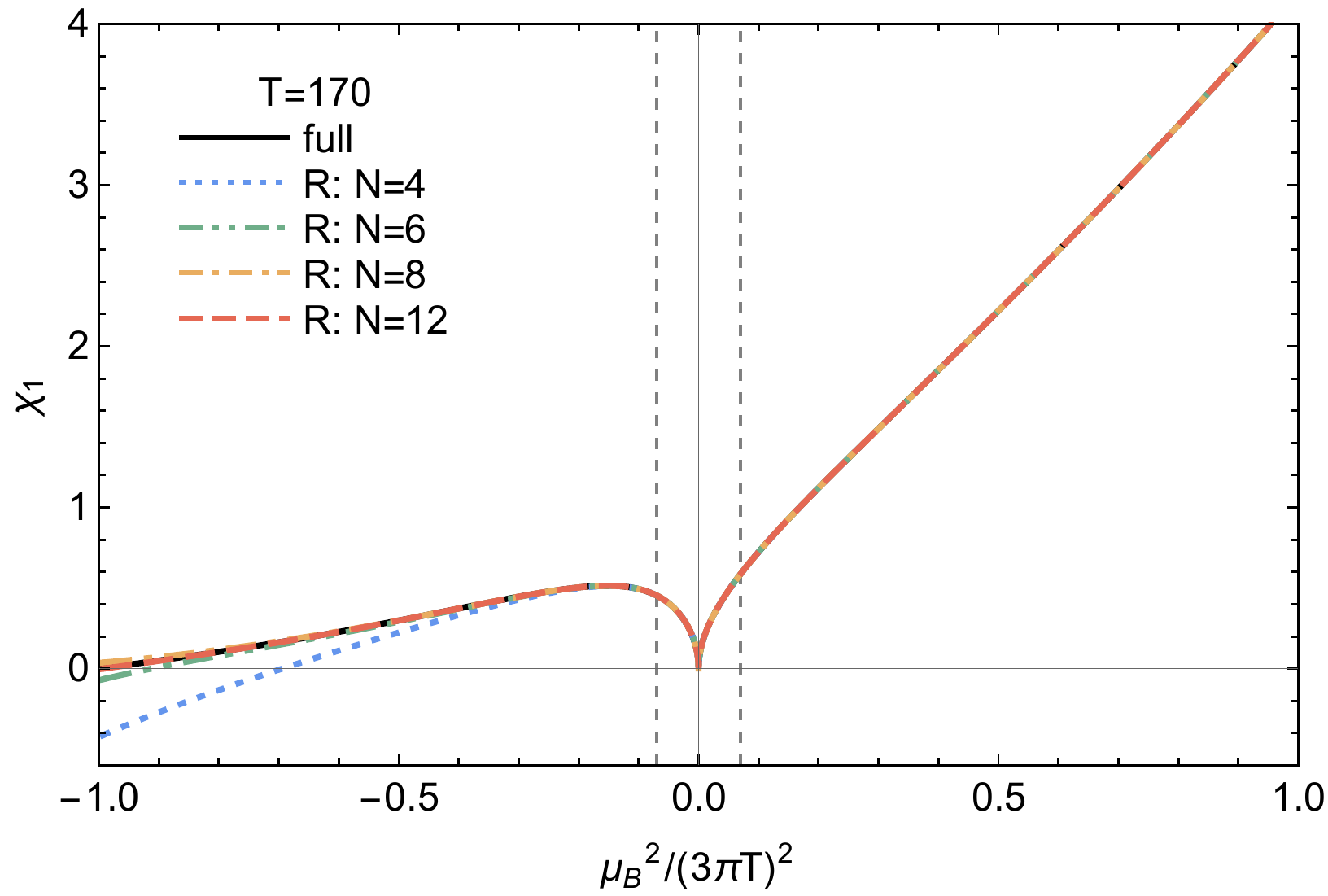}
\caption{$\chi_1(\mu_B^2)$ at $T=170\,{\rm MeV}$ for purely real and purely imaginary $\mu_B$. In the latter case, also $\chi_1$ is purely imaginary.
The vertical dashed lines denote the location of the edge singularity $|\mu_B^{\rm YL}|$.
{\it Left:} Comparison between the full result and various orders of the Taylor expansion around $\mu=0$ (E).
{\it Right:} Comparison between the full result and various orders of the resummation (R).
}
\label{fig:chi1T170}
\end{figure*}
%

%
\begin{figure*}[t]
\centering
\includegraphics[width=.48\textwidth]{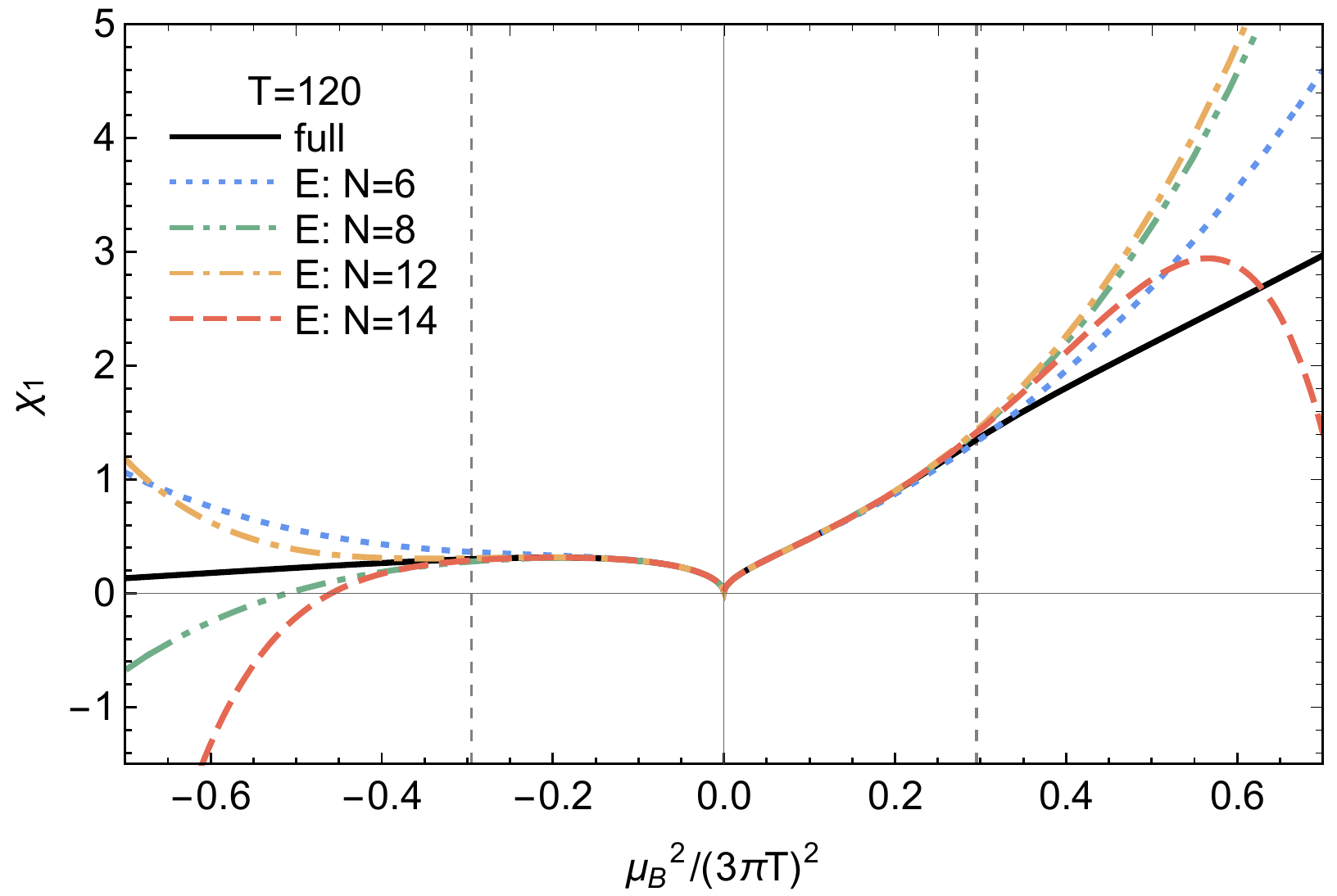}
\hfill
\includegraphics[width=.48\textwidth]{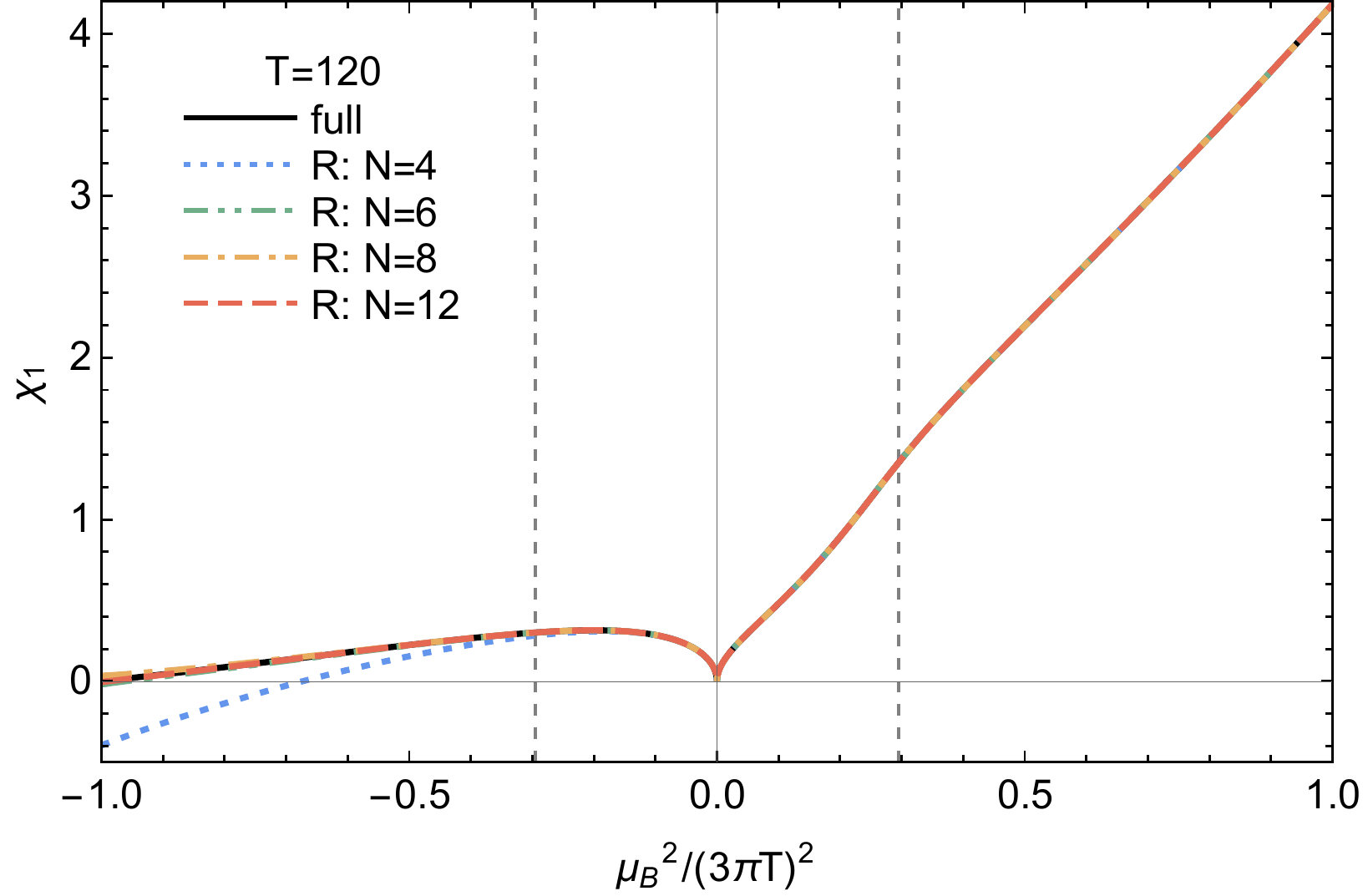}
\caption{Same as \Fig{fig:chi1T170} with $T=120$~MeV.}
\label{fig:chi1T120}
\end{figure*}
%

In addition to the singularity and cut associated to the Yang-Lee edge, there are also thermal cuts. These cuts basically follow from the analytical structure of \Eq{eq:JT} and are present even in a gas of free fermions. This can be seen from a small mass/$T$ expansion, which is valid in the symmetric phase,
\begin{align}\label{eq:highT}
\begin{split}
  J_{T,\mu_B}(\sigma) &= -\frac{1}{\pi} T^4\, {\rm Li}_4\big(-e^\frac{\mu_B}{3 T}\big)\\
  &\quad+ \frac{1}{16 \pi} h^2\sigma^2 T^2\, {\rm Li}_2\big(-e^\frac{\mu_B}{3 T}\big)  + \mathcal{O}\big(\sigma^4\big)\,.
\end{split}
\end{align}
The polylogarithm ${\rm Li}_s(z)$ has a branch cut at ${\rm Re}\, z > 1$ and ${\rm Im}\, z = 0$. This translates into cuts in the complex $\mu_B$ plane at ${\rm Re}\, \mu_B >0$ and ${\rm Im}\, \mu_B = 3 (2n+1)\pi T$ with $n \in \mathbb{Z}$. In both the Taylor expansion and the resummation this term is expanded in powers of $\mu_B$, so that these cuts cannot be resolved either way. We therefore limit our analysis to $|\mu_B| < 3 \pi T$, as it cannot be valid beyond this point. This also implies that the periodicity of $6 \pi T$ at purely imaginary $\mu_B$ of the quark-meson model cannot be captured by both schemes.
{In general, this follows from the fact that the power series expansion of ${\rm Li}_s(-e^w)$ about $w=0$ is only valid for $|w|<\pi$ \cite{dwood}.}

In our analysis $|\mu_B^{\rm YL}|>3 \pi T$ for $T\lesssim 80$~MeV. Thus, the thermal cut, rather than the the edge singularity, is the closest singularity at small temperatures. This explains why the determination of the edge singularity with the resummation scheme converges more slowly at smaller temperatures until it eventually fails when $T \leq |\mu_B^{\rm YL}(T)|/(3 \pi)$.  

We see that the program of \cite{Mondal:2021jxk} of using the resummation technique to locate Lee-Yang zeros (or at least the closest zeroes) 
finds support in our calculation. Note that our calculations are performed in the infinite volume/thermodynamic limit, while lattice QCD calculations are intrinsically finite volume. We remind the reader that, in the mean-field approximation, we are bound to consider the thermodynamic limit. 
In a finite volume the Yang-Lee edge and the corresponding branch cut will be replaced by set of Lee-Yang zeroes along 
the direction of cut.  The subtle difference between Lee-Yang zeroes and Yang-Lee edge is most probably of no consequence.

\section{\label{sec:thermo} Thermodynamics}

%
\begin{figure*}[t]
\centering
\includegraphics[width=.48\textwidth]{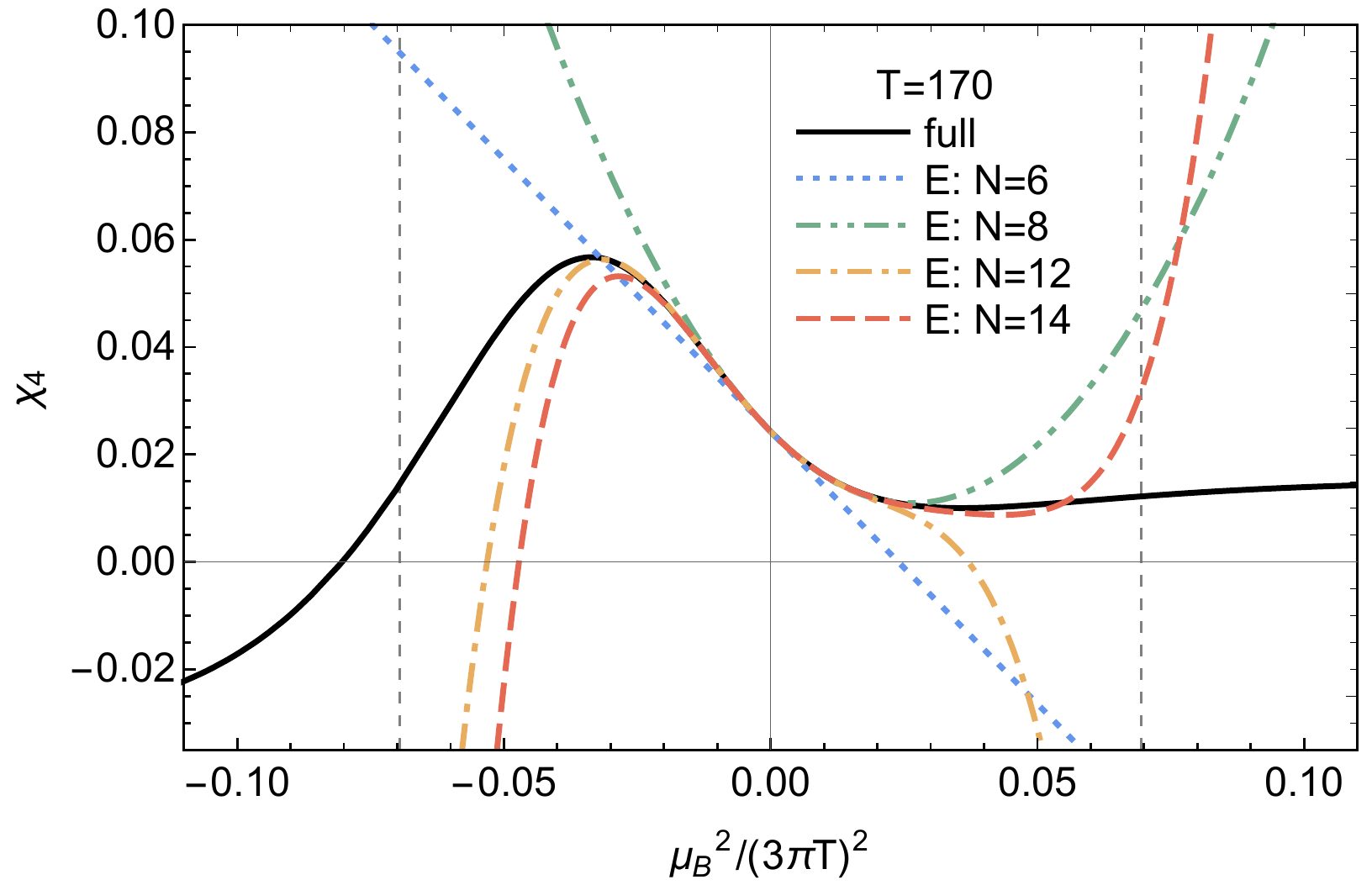}
\hfill
\includegraphics[width=.48\textwidth]{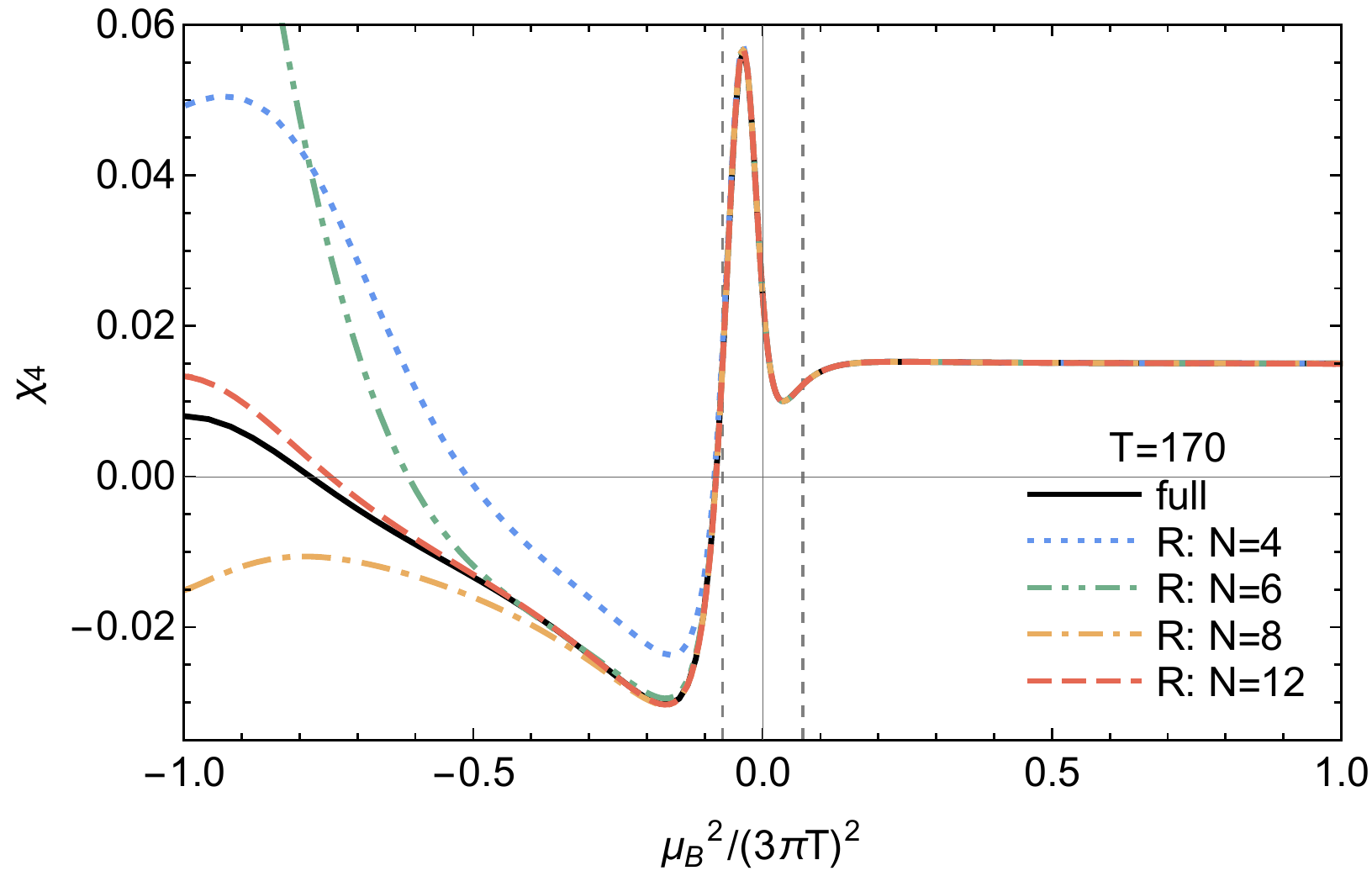}
\caption{$\chi_4(\mu_B^2)$ at $T=170\,{\rm MeV}$ for purely real and purely imaginary $\mu_B$.
The vertical dashed lines denote the location of the edge singularity $|\mu_B^{\rm YL}|$.
{\it Left:} Comparison between the full result and various orders of the Taylor expansion around $\mu=0$ (E).
{\it Right:} Comparison between the full result and various orders of the resummation (R).}
\label{fig:chi4T170}
\end{figure*}
%

%
\begin{figure*}[t]
\centering
\includegraphics[width=.48\textwidth]{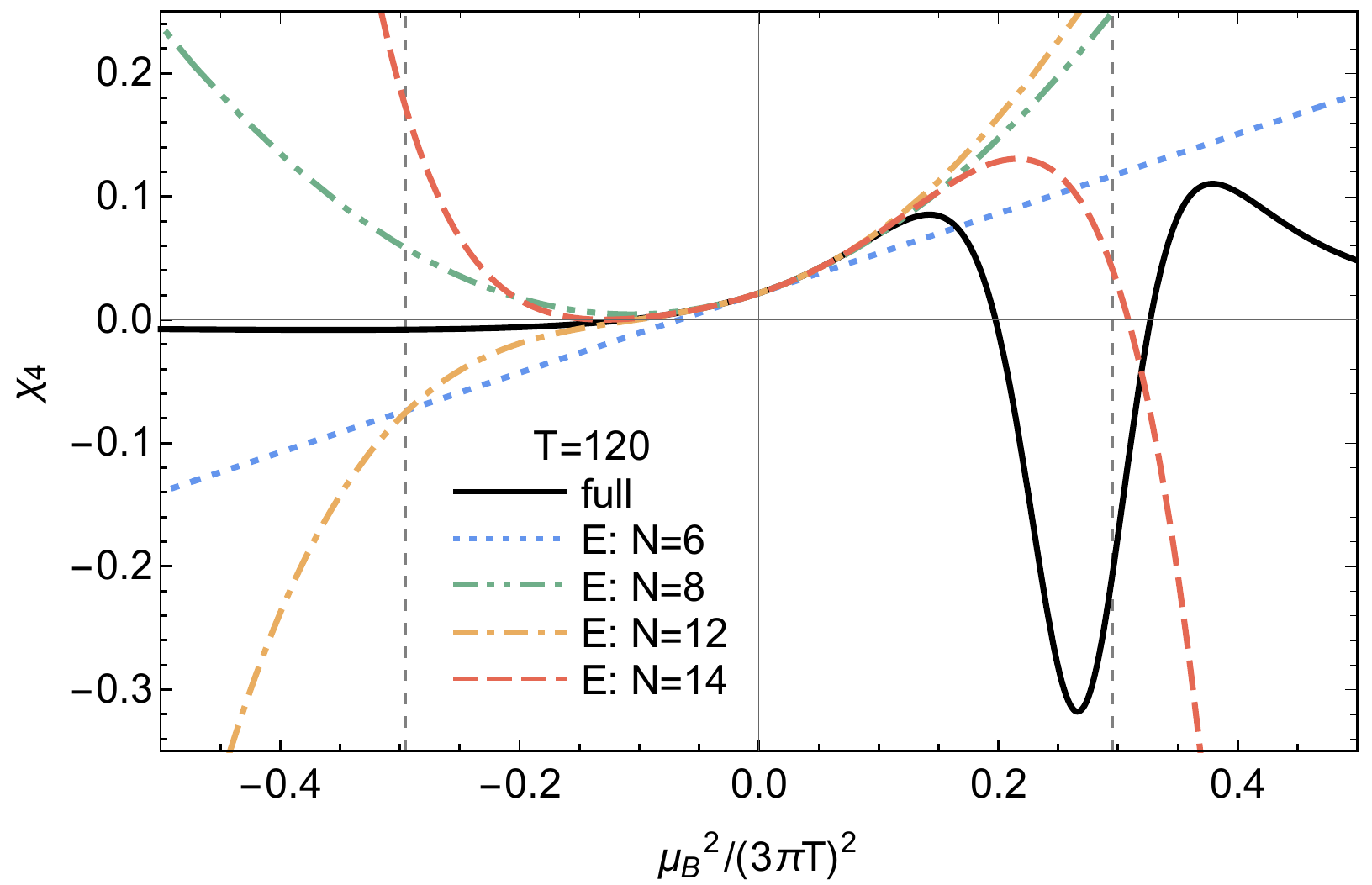}
\hfill
\includegraphics[width=.48\textwidth]{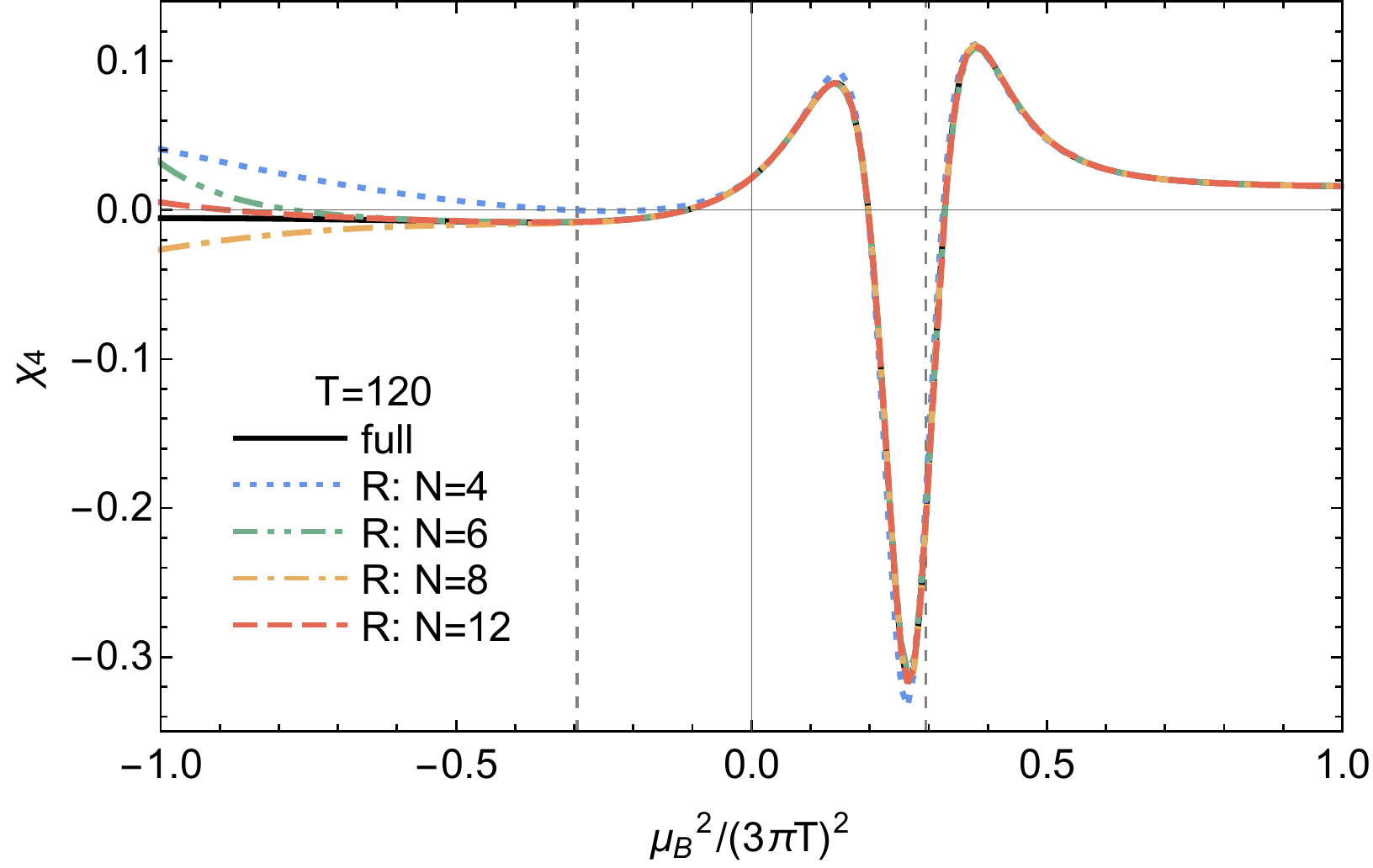}
\caption{Same as \Fig{fig:chi4T170} with $T=120$~MeV.}
\label{fig:chi4T120}
\end{figure*}
%

We now turn to thermodynamics of the model. Here we asses the accuracy of how well both the truncated Taylor series and the resummed approach reproduce the results obtained by a direct computation.    

In what follows we consider the net baryon density, $\chi_1$, and the fourth-order cumulant $\chi_4$ defined in \Eq{eq:sus}.
We choose these two quantities for the following reason. Although the baryon density is not very sensitive to the details of the equation of state, it is a key element in defining the equation of state of QCD and is required for the analysis of heavy-ion collision experiments. The fourth order cumulant is a more sensitive probe of the equation of state and is of significance for the experimental search of the critical end point.

{We consider two temperatures: one just a little bit above the pseudocritical temperature (see \Tab{tab:para}) and one below; $T=170$ and 120~MeV.
In Figs.~\ref{fig:chi1T170} and~\ref{fig:chi1T120} we show the dependence of the baryon density on real and imaginary chemical potential for these temperatures, respectively.} The left panels of these figures show the comparison between the direct calculations of $\chi_1$ with the results obtained by using the truncated Taylor series.
From both figures, it is evident that the Yang-Lee edge singularity, whose location $|\mu_B^{\rm YL}|$ is indicated by the vertical dashed gray lines,
limits the radius of the convergence of the series. Increasing the order of the truncation does not improve the 
convergence of the series for $|\mu_B| \gtrsim |\mu_B^{\rm YL}|$, as expected. This is true for both $T=170$~MeV (\Fig{fig:chi1T170}, left) and $T=120$~MeV (\Fig{fig:chi1T120}, left). The differences between both temperatures regarding the convergence behavior are only due to the different radii of convergence from the different locations of the edge singularity, cf.\ \Fig{fig:edge}.

The right panels of Figs.~\ref{fig:chi1T170} and~\ref{fig:chi1T120} show the comparison between the full result for $\chi_1$ and the resummation for different orders in the expansion of the fermion determinant. As compared to the Taylor expansion in the left panels of the respective figures, the resummation shows superior results without any apparent sensitivity to the radius of convergence defined by the Yang-Lee edge singularity.

The resummation converges rapidly with increasing order of the truncation within the whole region of real and imaginary chemical potentials studied here. $\chi_1$ is accurately described for all orders $N\geq 6$. The convergence at large imaginary chemical potential is slower than at large real chemical potential. This is because the chiral condensate $\bar\sigma$ monotonously increases from its value at $\mu_B = 0$ for increasing imaginary chemical potential with $i\mu_B \leq 3 i \pi T$, while it monotonously decreases with increasing real chemical potential. In the former case, the simple expansion that leads to \Eq{eq:highT} is not possible and the thermodynamic potential is in general a complicated function of $\mu_B$. In the latter case, for large real $\mu_B$, the condensate is very small and \Eq{eq:highT} is well approximated by only the first term. It then follows from the properties of the polylogarithm that the thermodynamic potential in \Eq{eq:barOmegaQM} is a fourth-order polynomial in $\mu_B$ for $|\mu_B|<3\pi T$ \cite{dwood}. Thus, for large real $\mu_B$, $\bar\sigma \rightarrow 0$ and the potential is that of a free gas of fermions. The resummation at order $N=4$ already captures this exactly. The Taylor expansion around $\mu_B=0$ is clearly not able to reproduce this simple asymptotic behavior at any order.

The dependence of the fourth order susceptibility on real and imaginary chemical potentials,  $\chi_4(\mu_B)$, is demonstrated in Figs.~\ref{fig:chi4T170} and~\ref{fig:chi4T120}, again at $T=170$ and 120 MeV. Since higher order susceptibilities are more sensitive to quark/baryon number fluctuations, $\chi_4$ is in general a more complicated function than $\chi_1$. It is therefore a stricter test for convergence. 
Furthermore, rapidly varying/increasing $\chi_n$ can be an indication for a crossover transition. This is seen in \Fig{fig:chi4T170}, where signs of a crossover at imaginary $\mu_B$ are shown, and similarly in \Fig{fig:chi4T120}, where a crossover is indicated at real $\mu_B$. This is expected because the former is at temperatures above $T_c$, while the latter is below. As can be seen in the left panels of these figures, the Taylor expansion is only converged for a very narrow region around $\mu_B = 0$ for $N\leq 14$. Thus, the order of the expansion is not high enough to probe the radius of convergence.

The convergence properties of the resummation discussed above for $\chi_1$, while qualitatively the same, are more apparent in $\chi_4$. This is shown in the right panels of Figs.~\ref{fig:chi4T170} and~\ref{fig:chi4T120}. Again, we observe very rapid convergence at real chemical potential. It follows from the previous discussion that $\chi_4$ has to become constant at large real $\mu_B$ when the condensate vanishes. This asymptotic behavior, as well as the nontrivial functional form of $\chi_4$ at smaller $\mu_B$, is reproduced with high precision already at low orders of the resummation. The convergence is markedly slower at large imaginary $\mu_B$, but still discernible. The range of validity of the resummation, $|\mu_B|<3\pi T$, discussed in \Sec{sec:ana}, is evident here.

\section{\label{sec:concl} Conclusions}

We have tested the scheme put forward in \cite{Mondal:2021jxk} for the resummation of infinite orders of an expansion of the partition function in the chemical potential in a mean-field quark-meson model. In this case, the resummation amounts to solving the equation of motion based on a series expansion of the effective potential. As a result, the bosonic mean-field becomes a nontrivial function of the chemical potential. This directly translates into a mean-field thermodynamic potential with nontrivial, in general nonanalytic, dependence on the chemical potential. This way, not only the effects of infinite powers in the chemical potential are taken into account, but also the analytic structure of the partition function becomes accessible.

This is in contrast to an ordinary Taylor expansion of the thermodynamic potential in powers of the chemical potential, as it is by definition strictly analytic and therefore bound to fail whenever nonanalyticities determine the structure of the partition function. The relevant singularities here are thermal cuts arising from thermal distributions in the partition function, and the Yang-Lee edge singularity, which is a critical point and a branch point singularity in the complex chemical potential plane.

By confronting the results of a direct computation at finite density with the resummation method and the Taylor expansion, we have tested the capabilities of describing the analytical structure and the thermodynamics of the model at finite density.
We have found that the resummation is far superior in describing the model at finite chemical potential as compared to the Taylor expansion. In fact, the resummation at truncation order $N=8$ already describes the susceptibilities for all real and imaginary chemical potentials studied here with high accuracy. Furthermore, the location of the Yang-Lee edge singularity is also described accurately for $T \gtrsim 90$~MeV.

Since the resummation is able to capture important analytical features of the equation of state at finite chemical potential, it is not limited by the same analytical constraints as the Taylor expansion. As expected, the Taylor expansion breaks down at the location of the edge singularity, since it defines its radius of convergence. Even high orders of the Taylor expansion fail to describe the nontrivial $\mu_B$-dependence of higher-order susceptibilities already at small $\mu_B$. In contrast, since infinite orders in $\mu_B$ are taken into account with the resummation, the nontrivial functional form of higher order susceptibilities, as well as their asymptotic behavior at large $\mu_B > 0$, is reproduced faithfully. We find that the resummation with increasing order of the truncation converges rapidly for chemical potentials in the region $-(3\pi T)^2 \leq \mu_B^2 \lesssim (3\pi T)^2$.
The only strict limit on the applicability of the resummation stems from the fact that the thermal cuts cannot be captured, limiting its range of validity to $|\mu_B|<3\pi T$ in the quark-meson model. In the confined phase of QCD this bound would be decreased to $|\mu_B|<\pi T$, as thermal distributions of baryons, rather than quarks, determine the thermal cuts in QCD.

We point out that the resummation is not aimed at curing or mitigating the sign problem; as for any reweighting method, \Eq{eq:ZR} has a sign problem. The goal is to capture the singularities in the complex plane in order to extend lattice computations to these values of $\mu_B$. The purpose of the present work has been to test whether or not the analytical structure can be captured by the resummation.

To further test the resummation scheme, a study analogous to the present one, but beyond mean-field, would be useful. In any case, the present work demonstrates the capability of this scheme in resolving the analytical structure and thermodynamics of a nontrivial theory for a wide range of complex chemical potentials. Our results therefore provide strong indications for the advantages of using the resummation scheme also for theories like QCD at finite baryochemical potential, as studied in \cite{Mondal:2021jxk}.

\section*{Acknowledgement}
We thank G.~Johnson and R.~Pisarski  for stimulating discussions. This material is based upon work supported by the U.S. Department of Energy, Office of Science, Office of Nuclear Physics through the Contract Nos. DE-SC0012704 (SM, FR), DE-SC0020081 (VS) and the Beam Energy Scan Theory (BEST) Topical Collaboration.       

\bibliography{resummation}

\begin{thebibliography}{35}%
\makeatletter
\providecommand \@ifxundefined [1]{%
 \@ifx{#1\undefined}
}%
\providecommand \@ifnum [1]{%
 \ifnum #1\expandafter \@firstoftwo
 \else \expandafter \@secondoftwo
 \fi
}%
\providecommand \@ifx [1]{%
 \ifx #1\expandafter \@firstoftwo
 \else \expandafter \@secondoftwo
 \fi
}%
\providecommand \natexlab [1]{#1}%
\providecommand \enquote  [1]{``#1''}%
\providecommand \bibnamefont  [1]{#1}%
\providecommand \bibfnamefont [1]{#1}%
\providecommand \citenamefont [1]{#1}%
\providecommand \href@noop [0]{\@secondoftwo}%
\providecommand \href [0]{\begingroup \@sanitize@url \@href}%
\providecommand \@href[1]{\@@startlink{#1}\@@href}%
\providecommand \@@href[1]{\endgroup#1\@@endlink}%
\providecommand \@sanitize@url [0]{\catcode `\\12\catcode `\$12\catcode
  `\&12\catcode `\#12\catcode `\^12\catcode `\_12\catcode `\%12\relax}%
\providecommand \@@startlink[1]{}%
\providecommand \@@endlink[0]{}%
\providecommand \url  [0]{\begingroup\@sanitize@url \@url }%
\providecommand \@url [1]{\endgroup\@href {#1}{\urlprefix }}%
\providecommand \urlprefix  [0]{URL }%
\providecommand \Eprint [0]{\href }%
\providecommand \doibase [0]{https://doi.org/}%
\providecommand \selectlanguage [0]{\@gobble}%
\providecommand \bibinfo  [0]{\@secondoftwo}%
\providecommand \bibfield  [0]{\@secondoftwo}%
\providecommand \translation [1]{[#1]}%
\providecommand \BibitemOpen [0]{}%
\providecommand \bibitemStop [0]{}%
\providecommand \bibitemNoStop [0]{.\EOS\space}%
\providecommand \EOS [0]{\spacefactor3000\relax}%
\providecommand \BibitemShut  [1]{\csname bibitem#1\endcsname}%
\let\auto@bib@innerbib\@empty
\bibitem [{\citenamefont {An}\ \emph {et~al.}(2021)\citenamefont {An} \emph
  {et~al.}}]{An:2021wof}%
  \BibitemOpen
  \bibfield  {author} {\bibinfo {author} {\bibfnamefont {X.}~\bibnamefont {An}}
  \emph {et~al.},\ }\bibfield  {title} {\bibinfo {title} {{The BEST framework
  for the search for the QCD critical point and the chiral magnetic effect}},\
  }\href@noop {} {\  (\bibinfo {year} {2021})},\ \Eprint
  {https://arxiv.org/abs/2108.13867} {arXiv:2108.13867 [nucl-th]} \BibitemShut
  {NoStop}%
\bibitem [{\citenamefont {Aarts}(2009)}]{Aarts:2009yj}%
  \BibitemOpen
  \bibfield  {author} {\bibinfo {author} {\bibfnamefont {G.}~\bibnamefont
  {Aarts}},\ }\bibfield  {title} {\bibinfo {title} {{Can complex Langevin
  dynamics evade the sign problem?}},\ }\href
  {https://doi.org/10.22323/1.091.0024} {\bibfield  {journal} {\bibinfo
  {journal} {PoS}\ }\textbf {\bibinfo {volume} {LAT2009}},\ \bibinfo {pages}
  {024} (\bibinfo {year} {2009})},\ \Eprint {https://arxiv.org/abs/0910.3772}
  {arXiv:0910.3772 [hep-lat]} \BibitemShut {NoStop}%
\bibitem [{\citenamefont {Aarts}\ \emph {et~al.}(2013)\citenamefont {Aarts},
  \citenamefont {Bongiovanni}, \citenamefont {Seiler}, \citenamefont {Sexty},\
  and\ \citenamefont {Stamatescu}}]{Aarts:2013uxa}%
  \BibitemOpen
  \bibfield  {author} {\bibinfo {author} {\bibfnamefont {G.}~\bibnamefont
  {Aarts}}, \bibinfo {author} {\bibfnamefont {L.}~\bibnamefont {Bongiovanni}},
  \bibinfo {author} {\bibfnamefont {E.}~\bibnamefont {Seiler}}, \bibinfo
  {author} {\bibfnamefont {D.}~\bibnamefont {Sexty}},\ and\ \bibinfo {author}
  {\bibfnamefont {I.-O.}\ \bibnamefont {Stamatescu}},\ }\bibfield  {title}
  {\bibinfo {title} {{Controlling complex Langevin dynamics at finite
  density}},\ }\href {https://doi.org/10.1140/epja/i2013-13089-4} {\bibfield
  {journal} {\bibinfo  {journal} {Eur. Phys. J. A}\ }\textbf {\bibinfo {volume}
  {49}},\ \bibinfo {pages} {89} (\bibinfo {year} {2013})},\ \Eprint
  {https://arxiv.org/abs/1303.6425} {arXiv:1303.6425 [hep-lat]} \BibitemShut
  {NoStop}%
\bibitem [{\citenamefont {Sexty}(2014)}]{Sexty:2013ica}%
  \BibitemOpen
  \bibfield  {author} {\bibinfo {author} {\bibfnamefont {D.}~\bibnamefont
  {Sexty}},\ }\bibfield  {title} {\bibinfo {title} {{Simulating full QCD at
  nonzero density using the complex Langevin equation}},\ }\href
  {https://doi.org/10.1016/j.physletb.2014.01.019} {\bibfield  {journal}
  {\bibinfo  {journal} {Phys. Lett. B}\ }\textbf {\bibinfo {volume} {729}},\
  \bibinfo {pages} {108} (\bibinfo {year} {2014})},\ \Eprint
  {https://arxiv.org/abs/1307.7748} {arXiv:1307.7748 [hep-lat]} \BibitemShut
  {NoStop}%
\bibitem [{\citenamefont {Fodor}\ \emph {et~al.}(2015)\citenamefont {Fodor},
  \citenamefont {Katz}, \citenamefont {Sexty},\ and\ \citenamefont
  {T\"or\"ok}}]{Fodor:2015doa}%
  \BibitemOpen
  \bibfield  {author} {\bibinfo {author} {\bibfnamefont {Z.}~\bibnamefont
  {Fodor}}, \bibinfo {author} {\bibfnamefont {S.~D.}\ \bibnamefont {Katz}},
  \bibinfo {author} {\bibfnamefont {D.}~\bibnamefont {Sexty}},\ and\ \bibinfo
  {author} {\bibfnamefont {C.}~\bibnamefont {T\"or\"ok}},\ }\bibfield  {title}
  {\bibinfo {title} {{Complex Langevin dynamics for dynamical QCD at nonzero
  chemical potential: A comparison with multiparameter reweighting}},\ }\href
  {https://doi.org/10.1103/PhysRevD.92.094516} {\bibfield  {journal} {\bibinfo
  {journal} {Phys. Rev. D}\ }\textbf {\bibinfo {volume} {92}},\ \bibinfo
  {pages} {094516} (\bibinfo {year} {2015})},\ \Eprint
  {https://arxiv.org/abs/1508.05260} {arXiv:1508.05260 [hep-lat]} \BibitemShut
  {NoStop}%
\bibitem [{\citenamefont {Cristoforetti}\ \emph {et~al.}(2012)\citenamefont
  {Cristoforetti}, \citenamefont {Di~Renzo},\ and\ \citenamefont
  {Scorzato}}]{Cristoforetti:2012su}%
  \BibitemOpen
  \bibfield  {author} {\bibinfo {author} {\bibfnamefont {M.}~\bibnamefont
  {Cristoforetti}}, \bibinfo {author} {\bibfnamefont {F.}~\bibnamefont
  {Di~Renzo}},\ and\ \bibinfo {author} {\bibfnamefont {L.}~\bibnamefont
  {Scorzato}} (\bibinfo {collaboration} {AuroraScience}),\ }\bibfield  {title}
  {\bibinfo {title} {{New approach to the sign problem in quantum field
  theories: High density QCD on a Lefschetz thimble}},\ }\href
  {https://doi.org/10.1103/PhysRevD.86.074506} {\bibfield  {journal} {\bibinfo
  {journal} {Phys. Rev. D}\ }\textbf {\bibinfo {volume} {86}},\ \bibinfo
  {pages} {074506} (\bibinfo {year} {2012})},\ \Eprint
  {https://arxiv.org/abs/1205.3996} {arXiv:1205.3996 [hep-lat]} \BibitemShut
  {NoStop}%
\bibitem [{\citenamefont {Fukuma}\ \emph {et~al.}(2019)\citenamefont {Fukuma},
  \citenamefont {Matsumoto},\ and\ \citenamefont {Umeda}}]{Fukuma:2019uot}%
  \BibitemOpen
  \bibfield  {author} {\bibinfo {author} {\bibfnamefont {M.}~\bibnamefont
  {Fukuma}}, \bibinfo {author} {\bibfnamefont {N.}~\bibnamefont {Matsumoto}},\
  and\ \bibinfo {author} {\bibfnamefont {N.}~\bibnamefont {Umeda}},\ }\bibfield
   {title} {\bibinfo {title} {{Implementation of the HMC algorithm on the
  tempered Lefschetz thimble method}},\ }\href@noop {} {\  (\bibinfo {year}
  {2019})},\ \Eprint {https://arxiv.org/abs/1912.13303} {arXiv:1912.13303
  [hep-lat]} \BibitemShut {NoStop}%
\bibitem [{\citenamefont {Alexandru}\ \emph {et~al.}(2020)\citenamefont
  {Alexandru}, \citenamefont {Basar}, \citenamefont {Bedaque},\ and\
  \citenamefont {Warrington}}]{Alexandru:2020wrj}%
  \BibitemOpen
  \bibfield  {author} {\bibinfo {author} {\bibfnamefont {A.}~\bibnamefont
  {Alexandru}}, \bibinfo {author} {\bibfnamefont {G.}~\bibnamefont {Basar}},
  \bibinfo {author} {\bibfnamefont {P.~F.}\ \bibnamefont {Bedaque}},\ and\
  \bibinfo {author} {\bibfnamefont {N.~C.}\ \bibnamefont {Warrington}},\
  }\bibfield  {title} {\bibinfo {title} {{Complex Paths Around The Sign
  Problem}},\ }\href@noop {} {\  (\bibinfo {year} {2020})},\ \Eprint
  {https://arxiv.org/abs/2007.05436} {arXiv:2007.05436 [hep-lat]} \BibitemShut
  {NoStop}%
\bibitem [{\citenamefont {Fischer}(2019)}]{Fischer:2018sdj}%
  \BibitemOpen
  \bibfield  {author} {\bibinfo {author} {\bibfnamefont {C.~S.}\ \bibnamefont
  {Fischer}},\ }\bibfield  {title} {\bibinfo {title} {{QCD at finite
  temperature and chemical potential from Dyson\textendash{}Schwinger
  equations}},\ }\href {https://doi.org/10.1016/j.ppnp.2019.01.002} {\bibfield
  {journal} {\bibinfo  {journal} {Prog. Part. Nucl. Phys.}\ }\textbf {\bibinfo
  {volume} {105}},\ \bibinfo {pages} {1} (\bibinfo {year} {2019})},\ \Eprint
  {https://arxiv.org/abs/1810.12938} {arXiv:1810.12938 [hep-ph]} \BibitemShut
  {NoStop}%
\bibitem [{\citenamefont {Fu}\ \emph {et~al.}(2020)\citenamefont {Fu},
  \citenamefont {Pawlowski},\ and\ \citenamefont {Rennecke}}]{Fu:2019hdw}%
  \BibitemOpen
  \bibfield  {author} {\bibinfo {author} {\bibfnamefont {W.-j.}\ \bibnamefont
  {Fu}}, \bibinfo {author} {\bibfnamefont {J.~M.}\ \bibnamefont {Pawlowski}},\
  and\ \bibinfo {author} {\bibfnamefont {F.}~\bibnamefont {Rennecke}},\
  }\bibfield  {title} {\bibinfo {title} {{QCD phase structure at finite
  temperature and density}},\ }\href
  {https://doi.org/10.1103/PhysRevD.101.054032} {\bibfield  {journal} {\bibinfo
   {journal} {Phys. Rev. D}\ }\textbf {\bibinfo {volume} {101}},\ \bibinfo
  {pages} {054032} (\bibinfo {year} {2020})},\ \Eprint
  {https://arxiv.org/abs/1909.02991} {arXiv:1909.02991 [hep-ph]} \BibitemShut
  {NoStop}%
\bibitem [{\citenamefont {Bazavov}\ \emph {et~al.}(2017)\citenamefont {Bazavov}
  \emph {et~al.}}]{Bazavov:2017dus}%
  \BibitemOpen
  \bibfield  {author} {\bibinfo {author} {\bibfnamefont {A.}~\bibnamefont
  {Bazavov}} \emph {et~al.},\ }\bibfield  {title} {\bibinfo {title} {{The QCD
  Equation of State to $\mathcal{O}(\mu_B^6)$ from Lattice QCD}},\ }\href
  {https://doi.org/10.1103/PhysRevD.95.054504} {\bibfield  {journal} {\bibinfo
  {journal} {Phys. Rev. D}\ }\textbf {\bibinfo {volume} {95}},\ \bibinfo
  {pages} {054504} (\bibinfo {year} {2017})},\ \Eprint
  {https://arxiv.org/abs/1701.04325} {arXiv:1701.04325 [hep-lat]} \BibitemShut
  {NoStop}%
\bibitem [{\citenamefont {Datta}\ \emph {et~al.}(2017)\citenamefont {Datta},
  \citenamefont {Gavai},\ and\ \citenamefont {Gupta}}]{Datta:2016ukp}%
  \BibitemOpen
  \bibfield  {author} {\bibinfo {author} {\bibfnamefont {S.}~\bibnamefont
  {Datta}}, \bibinfo {author} {\bibfnamefont {R.~V.}\ \bibnamefont {Gavai}},\
  and\ \bibinfo {author} {\bibfnamefont {S.}~\bibnamefont {Gupta}},\ }\bibfield
   {title} {\bibinfo {title} {{Quark number susceptibilities and equation of
  state at finite chemical potential in staggered QCD with Nt=8}},\ }\href
  {https://doi.org/10.1103/PhysRevD.95.054512} {\bibfield  {journal} {\bibinfo
  {journal} {Phys. Rev. D}\ }\textbf {\bibinfo {volume} {95}},\ \bibinfo
  {pages} {054512} (\bibinfo {year} {2017})},\ \Eprint
  {https://arxiv.org/abs/1612.06673} {arXiv:1612.06673 [hep-lat]} \BibitemShut
  {NoStop}%
\bibitem [{\citenamefont {Bors\'anyi}\ \emph {et~al.}(2021)\citenamefont
  {Bors\'anyi}, \citenamefont {Fodor}, \citenamefont {Guenther}, \citenamefont
  {Kara}, \citenamefont {Katz}, \citenamefont {Parotto}, \citenamefont
  {P\'asztor}, \citenamefont {Ratti},\ and\ \citenamefont
  {Szab\'o}}]{Borsanyi:2021sxv}%
  \BibitemOpen
  \bibfield  {author} {\bibinfo {author} {\bibfnamefont {S.}~\bibnamefont
  {Bors\'anyi}}, \bibinfo {author} {\bibfnamefont {Z.}~\bibnamefont {Fodor}},
  \bibinfo {author} {\bibfnamefont {J.~N.}\ \bibnamefont {Guenther}}, \bibinfo
  {author} {\bibfnamefont {R.}~\bibnamefont {Kara}}, \bibinfo {author}
  {\bibfnamefont {S.~D.}\ \bibnamefont {Katz}}, \bibinfo {author}
  {\bibfnamefont {P.}~\bibnamefont {Parotto}}, \bibinfo {author} {\bibfnamefont
  {A.}~\bibnamefont {P\'asztor}}, \bibinfo {author} {\bibfnamefont
  {C.}~\bibnamefont {Ratti}},\ and\ \bibinfo {author} {\bibfnamefont {K.~K.}\
  \bibnamefont {Szab\'o}},\ }\bibfield  {title} {\bibinfo {title} {{Lattice QCD
  equation of state at finite chemical potential from an alternative expansion
  scheme}},\ }\href {https://doi.org/10.1103/PhysRevLett.126.232001} {\bibfield
   {journal} {\bibinfo  {journal} {Phys. Rev. Lett.}\ }\textbf {\bibinfo
  {volume} {126}},\ \bibinfo {pages} {232001} (\bibinfo {year} {2021})},\
  \Eprint {https://arxiv.org/abs/2102.06660} {arXiv:2102.06660 [hep-lat]}
  \BibitemShut {NoStop}%
\bibitem [{\citenamefont {Gavai}\ and\ \citenamefont
  {Gupta}(2008)}]{Gavai:2008zr}%
  \BibitemOpen
  \bibfield  {author} {\bibinfo {author} {\bibfnamefont {R.~V.}\ \bibnamefont
  {Gavai}}\ and\ \bibinfo {author} {\bibfnamefont {S.}~\bibnamefont {Gupta}},\
  }\bibfield  {title} {\bibinfo {title} {{QCD at finite chemical potential with
  six time slices}},\ }\href {https://doi.org/10.1103/PhysRevD.78.114503}
  {\bibfield  {journal} {\bibinfo  {journal} {Phys. Rev. D}\ }\textbf {\bibinfo
  {volume} {78}},\ \bibinfo {pages} {114503} (\bibinfo {year} {2008})},\
  \Eprint {https://arxiv.org/abs/0806.2233} {arXiv:0806.2233 [hep-lat]}
  \BibitemShut {NoStop}%
\bibitem [{\citenamefont {Karsch}\ \emph {et~al.}(2011)\citenamefont {Karsch},
  \citenamefont {Schaefer}, \citenamefont {Wagner},\ and\ \citenamefont
  {Wambach}}]{Karsch:2010hm}%
  \BibitemOpen
  \bibfield  {author} {\bibinfo {author} {\bibfnamefont {F.}~\bibnamefont
  {Karsch}}, \bibinfo {author} {\bibfnamefont {B.-J.}\ \bibnamefont
  {Schaefer}}, \bibinfo {author} {\bibfnamefont {M.}~\bibnamefont {Wagner}},\
  and\ \bibinfo {author} {\bibfnamefont {J.}~\bibnamefont {Wambach}},\
  }\bibfield  {title} {\bibinfo {title} {{Towards finite density QCD with
  Taylor expansions}},\ }\href {https://doi.org/10.1016/j.physletb.2011.03.013}
  {\bibfield  {journal} {\bibinfo  {journal} {Phys. Lett. B}\ }\textbf
  {\bibinfo {volume} {698}},\ \bibinfo {pages} {256} (\bibinfo {year}
  {2011})},\ \Eprint {https://arxiv.org/abs/1009.5211} {arXiv:1009.5211
  [hep-ph]} \BibitemShut {NoStop}%
\bibitem [{\citenamefont {Basar}(2021)}]{Basar:2021hdf}%
  \BibitemOpen
  \bibfield  {author} {\bibinfo {author} {\bibfnamefont {G.}~\bibnamefont
  {Basar}},\ }\bibfield  {title} {\bibinfo {title} {{Universality, Lee-Yang
  Singularities, and Series Expansions}},\ }\href
  {https://doi.org/10.1103/PhysRevLett.127.171603} {\bibfield  {journal}
  {\bibinfo  {journal} {Phys. Rev. Lett.}\ }\textbf {\bibinfo {volume} {127}},\
  \bibinfo {pages} {171603} (\bibinfo {year} {2021})},\ \Eprint
  {https://arxiv.org/abs/2105.08080} {arXiv:2105.08080 [hep-th]} \BibitemShut
  {NoStop}%
\bibitem [{\citenamefont {Schmidt}\ \emph {et~al.}(2021)\citenamefont
  {Schmidt}, \citenamefont {Goswami}, \citenamefont {Nicotra}, \citenamefont
  {Ziesch\'e}, \citenamefont {Dimopoulos}, \citenamefont {Di~Renzo},
  \citenamefont {Singh},\ and\ \citenamefont {Zambello}}]{Schmidt:2021pey}%
  \BibitemOpen
  \bibfield  {author} {\bibinfo {author} {\bibfnamefont {C.}~\bibnamefont
  {Schmidt}}, \bibinfo {author} {\bibfnamefont {J.}~\bibnamefont {Goswami}},
  \bibinfo {author} {\bibfnamefont {G.}~\bibnamefont {Nicotra}}, \bibinfo
  {author} {\bibfnamefont {F.}~\bibnamefont {Ziesch\'e}}, \bibinfo {author}
  {\bibfnamefont {P.}~\bibnamefont {Dimopoulos}}, \bibinfo {author}
  {\bibfnamefont {F.}~\bibnamefont {Di~Renzo}}, \bibinfo {author}
  {\bibfnamefont {S.}~\bibnamefont {Singh}},\ and\ \bibinfo {author}
  {\bibfnamefont {K.}~\bibnamefont {Zambello}},\ }\bibfield  {title} {\bibinfo
  {title} {{Net-baryon number fluctuations}},\ }in\ \href@noop {} {\emph
  {\bibinfo {booktitle} {{Criticality in QCD and the Hadron Resonance Gas}}}}\
  (\bibinfo {year} {2021})\ \Eprint {https://arxiv.org/abs/2101.02254}
  {arXiv:2101.02254 [hep-lat]} \BibitemShut {NoStop}%
\bibitem [{\citenamefont {Mondal}\ \emph {et~al.}(2022)\citenamefont {Mondal},
  \citenamefont {Mukherjee},\ and\ \citenamefont {Hegde}}]{Mondal:2021jxk}%
  \BibitemOpen
  \bibfield  {author} {\bibinfo {author} {\bibfnamefont {S.}~\bibnamefont
  {Mondal}}, \bibinfo {author} {\bibfnamefont {S.}~\bibnamefont {Mukherjee}},\
  and\ \bibinfo {author} {\bibfnamefont {P.}~\bibnamefont {Hegde}},\ }\bibfield
   {title} {\bibinfo {title} {{Lattice QCD Equation of State for Nonvanishing
  Chemical Potential by Resumming Taylor Expansions}},\ }\href
  {https://doi.org/10.1103/PhysRevLett.128.022001} {\bibfield  {journal}
  {\bibinfo  {journal} {Phys. Rev. Lett.}\ }\textbf {\bibinfo {volume} {128}},\
  \bibinfo {pages} {022001} (\bibinfo {year} {2022})},\ \Eprint
  {https://arxiv.org/abs/2106.03165} {arXiv:2106.03165 [hep-lat]} \BibitemShut
  {NoStop}%
\bibitem [{\citenamefont {Jungnickel}\ and\ \citenamefont
  {Wetterich}(1996)}]{Jungnickel:1995fp}%
  \BibitemOpen
  \bibfield  {author} {\bibinfo {author} {\bibfnamefont {D.~U.}\ \bibnamefont
  {Jungnickel}}\ and\ \bibinfo {author} {\bibfnamefont {C.}~\bibnamefont
  {Wetterich}},\ }\bibfield  {title} {\bibinfo {title} {{Effective action for
  the chiral quark-meson model}},\ }\href
  {https://doi.org/10.1103/PhysRevD.53.5142} {\bibfield  {journal} {\bibinfo
  {journal} {Phys. Rev. D}\ }\textbf {\bibinfo {volume} {53}},\ \bibinfo
  {pages} {5142} (\bibinfo {year} {1996})},\ \Eprint
  {https://arxiv.org/abs/hep-ph/9505267} {arXiv:hep-ph/9505267} \BibitemShut
  {NoStop}%
\bibitem [{\citenamefont {Schaefer}\ and\ \citenamefont
  {Wambach}(2005)}]{Schaefer:2004en}%
  \BibitemOpen
  \bibfield  {author} {\bibinfo {author} {\bibfnamefont {B.-J.}\ \bibnamefont
  {Schaefer}}\ and\ \bibinfo {author} {\bibfnamefont {J.}~\bibnamefont
  {Wambach}},\ }\bibfield  {title} {\bibinfo {title} {{The Phase diagram of the
  quark meson model}},\ }\href
  {https://doi.org/10.1016/j.nuclphysa.2005.04.012} {\bibfield  {journal}
  {\bibinfo  {journal} {Nucl. Phys. A}\ }\textbf {\bibinfo {volume} {757}},\
  \bibinfo {pages} {479} (\bibinfo {year} {2005})},\ \Eprint
  {https://arxiv.org/abs/nucl-th/0403039} {arXiv:nucl-th/0403039} \BibitemShut
  {NoStop}%
\bibitem [{\citenamefont {Skokov}\ \emph {et~al.}(2010)\citenamefont {Skokov},
  \citenamefont {Friman}, \citenamefont {Nakano}, \citenamefont {Redlich},\
  and\ \citenamefont {Schaefer}}]{Skokov:2010sf}%
  \BibitemOpen
  \bibfield  {author} {\bibinfo {author} {\bibfnamefont {V.}~\bibnamefont
  {Skokov}}, \bibinfo {author} {\bibfnamefont {B.}~\bibnamefont {Friman}},
  \bibinfo {author} {\bibfnamefont {E.}~\bibnamefont {Nakano}}, \bibinfo
  {author} {\bibfnamefont {K.}~\bibnamefont {Redlich}},\ and\ \bibinfo {author}
  {\bibfnamefont {B.~J.}\ \bibnamefont {Schaefer}},\ }\bibfield  {title}
  {\bibinfo {title} {{Vacuum fluctuations and the thermodynamics of chiral
  models}},\ }\href {https://doi.org/10.1103/PhysRevD.82.034029} {\bibfield
  {journal} {\bibinfo  {journal} {Phys. Rev. D}\ }\textbf {\bibinfo {volume}
  {82}},\ \bibinfo {pages} {034029} (\bibinfo {year} {2010})},\ \Eprint
  {https://arxiv.org/abs/1005.3166} {arXiv:1005.3166 [hep-ph]} \BibitemShut
  {NoStop}%
\bibitem [{\citenamefont {Pawlowski}\ and\ \citenamefont
  {Rennecke}(2014)}]{Pawlowski:2014zaa}%
  \BibitemOpen
  \bibfield  {author} {\bibinfo {author} {\bibfnamefont {J.~M.}\ \bibnamefont
  {Pawlowski}}\ and\ \bibinfo {author} {\bibfnamefont {F.}~\bibnamefont
  {Rennecke}},\ }\bibfield  {title} {\bibinfo {title} {{Higher order
  quark-mesonic scattering processes and the phase structure of QCD}},\ }\href
  {https://doi.org/10.1103/PhysRevD.90.076002} {\bibfield  {journal} {\bibinfo
  {journal} {Phys. Rev. D}\ }\textbf {\bibinfo {volume} {90}},\ \bibinfo
  {pages} {076002} (\bibinfo {year} {2014})},\ \Eprint
  {https://arxiv.org/abs/1403.1179} {arXiv:1403.1179 [hep-ph]} \BibitemShut
  {NoStop}%
\bibitem [{\citenamefont {Resch}\ \emph {et~al.}(2019)\citenamefont {Resch},
  \citenamefont {Rennecke},\ and\ \citenamefont {Schaefer}}]{Resch:2017vjs}%
  \BibitemOpen
  \bibfield  {author} {\bibinfo {author} {\bibfnamefont {S.}~\bibnamefont
  {Resch}}, \bibinfo {author} {\bibfnamefont {F.}~\bibnamefont {Rennecke}},\
  and\ \bibinfo {author} {\bibfnamefont {B.-J.}\ \bibnamefont {Schaefer}},\
  }\bibfield  {title} {\bibinfo {title} {{Mass sensitivity of the three-flavor
  chiral phase transition}},\ }\href
  {https://doi.org/10.1103/PhysRevD.99.076005} {\bibfield  {journal} {\bibinfo
  {journal} {Phys. Rev. D}\ }\textbf {\bibinfo {volume} {99}},\ \bibinfo
  {pages} {076005} (\bibinfo {year} {2019})},\ \Eprint
  {https://arxiv.org/abs/1712.07961} {arXiv:1712.07961 [hep-ph]} \BibitemShut
  {NoStop}%
\bibitem [{\citenamefont {Allton}\ \emph {et~al.}(2005)\citenamefont {Allton},
  \citenamefont {Doring}, \citenamefont {Ejiri}, \citenamefont {Hands},
  \citenamefont {Kaczmarek}, \citenamefont {Karsch}, \citenamefont {Laermann},\
  and\ \citenamefont {Redlich}}]{Allton:2005gk}%
  \BibitemOpen
  \bibfield  {author} {\bibinfo {author} {\bibfnamefont {C.~R.}\ \bibnamefont
  {Allton}}, \bibinfo {author} {\bibfnamefont {M.}~\bibnamefont {Doring}},
  \bibinfo {author} {\bibfnamefont {S.}~\bibnamefont {Ejiri}}, \bibinfo
  {author} {\bibfnamefont {S.~J.}\ \bibnamefont {Hands}}, \bibinfo {author}
  {\bibfnamefont {O.}~\bibnamefont {Kaczmarek}}, \bibinfo {author}
  {\bibfnamefont {F.}~\bibnamefont {Karsch}}, \bibinfo {author} {\bibfnamefont
  {E.}~\bibnamefont {Laermann}},\ and\ \bibinfo {author} {\bibfnamefont
  {K.}~\bibnamefont {Redlich}},\ }\bibfield  {title} {\bibinfo {title}
  {{Thermodynamics of two flavor QCD to sixth order in quark chemical
  potential}},\ }\href {https://doi.org/10.1103/PhysRevD.71.054508} {\bibfield
  {journal} {\bibinfo  {journal} {Phys. Rev. D}\ }\textbf {\bibinfo {volume}
  {71}},\ \bibinfo {pages} {054508} (\bibinfo {year} {2005})},\ \Eprint
  {https://arxiv.org/abs/hep-lat/0501030} {arXiv:hep-lat/0501030} \BibitemShut
  {NoStop}%
\bibitem [{\citenamefont {Giordano}\ \emph {et~al.}(2020)\citenamefont
  {Giordano}, \citenamefont {Kapas}, \citenamefont {Katz}, \citenamefont
  {Nogradi},\ and\ \citenamefont {Pasztor}}]{Giordano:2020roi}%
  \BibitemOpen
  \bibfield  {author} {\bibinfo {author} {\bibfnamefont {M.}~\bibnamefont
  {Giordano}}, \bibinfo {author} {\bibfnamefont {K.}~\bibnamefont {Kapas}},
  \bibinfo {author} {\bibfnamefont {S.~D.}\ \bibnamefont {Katz}}, \bibinfo
  {author} {\bibfnamefont {D.}~\bibnamefont {Nogradi}},\ and\ \bibinfo {author}
  {\bibfnamefont {A.}~\bibnamefont {Pasztor}},\ }\bibfield  {title} {\bibinfo
  {title} {{New approach to lattice QCD at finite density; results for the
  critical end point on coarse lattices}},\ }\href
  {https://doi.org/10.1007/JHEP05(2020)088} {\bibfield  {journal} {\bibinfo
  {journal} {JHEP}\ }\textbf {\bibinfo {volume} {05}},\ \bibinfo {pages}
  {088}},\ \Eprint {https://arxiv.org/abs/2004.10800} {arXiv:2004.10800
  [hep-lat]} \BibitemShut {NoStop}%
\bibitem [{\citenamefont {Borsanyi}\ \emph {et~al.}(2022)\citenamefont
  {Borsanyi}, \citenamefont {Fodor}, \citenamefont {Giordano}, \citenamefont
  {Katz}, \citenamefont {Nogradi}, \citenamefont {Pasztor},\ and\ \citenamefont
  {Wong}}]{Borsanyi:2021hbk}%
  \BibitemOpen
  \bibfield  {author} {\bibinfo {author} {\bibfnamefont {S.}~\bibnamefont
  {Borsanyi}}, \bibinfo {author} {\bibfnamefont {Z.}~\bibnamefont {Fodor}},
  \bibinfo {author} {\bibfnamefont {M.}~\bibnamefont {Giordano}}, \bibinfo
  {author} {\bibfnamefont {S.~D.}\ \bibnamefont {Katz}}, \bibinfo {author}
  {\bibfnamefont {D.}~\bibnamefont {Nogradi}}, \bibinfo {author} {\bibfnamefont
  {A.}~\bibnamefont {Pasztor}},\ and\ \bibinfo {author} {\bibfnamefont {C.~H.}\
  \bibnamefont {Wong}},\ }\bibfield  {title} {\bibinfo {title} {{Lattice
  simulations of the QCD chiral transition at real baryon density}},\ }\href
  {https://doi.org/10.1103/PhysRevD.105.L051506} {\bibfield  {journal}
  {\bibinfo  {journal} {Phys. Rev. D}\ }\textbf {\bibinfo {volume} {105}},\
  \bibinfo {pages} {L051506} (\bibinfo {year} {2022})},\ \Eprint
  {https://arxiv.org/abs/2108.09213} {arXiv:2108.09213 [hep-lat]} \BibitemShut
  {NoStop}%
\bibitem [{\citenamefont {Yang}\ and\ \citenamefont {Lee}(1952)}]{Yang:1952be}%
  \BibitemOpen
  \bibfield  {author} {\bibinfo {author} {\bibfnamefont {C.-N.}\ \bibnamefont
  {Yang}}\ and\ \bibinfo {author} {\bibfnamefont {T.~D.}\ \bibnamefont {Lee}},\
  }\bibfield  {title} {\bibinfo {title} {{Statistical theory of equations of
  state and phase transitions. 1. Theory of condensation}},\ }\href
  {https://doi.org/10.1103/PhysRev.87.404} {\bibfield  {journal} {\bibinfo
  {journal} {Phys. Rev.}\ }\textbf {\bibinfo {volume} {87}},\ \bibinfo {pages}
  {404} (\bibinfo {year} {1952})}\BibitemShut {NoStop}%
\bibitem [{\citenamefont {Lee}\ and\ \citenamefont {Yang}(1952)}]{Lee:1952ig}%
  \BibitemOpen
  \bibfield  {author} {\bibinfo {author} {\bibfnamefont {T.~D.}\ \bibnamefont
  {Lee}}\ and\ \bibinfo {author} {\bibfnamefont {C.-N.}\ \bibnamefont {Yang}},\
  }\bibfield  {title} {\bibinfo {title} {{Statistical theory of equations of
  state and phase transitions. 2. Lattice gas and Ising model}},\ }\href
  {https://doi.org/10.1103/PhysRev.87.410} {\bibfield  {journal} {\bibinfo
  {journal} {Phys. Rev.}\ }\textbf {\bibinfo {volume} {87}},\ \bibinfo {pages}
  {410} (\bibinfo {year} {1952})}\BibitemShut {NoStop}%
\bibitem [{\citenamefont {Laine}\ and\ \citenamefont
  {Vuorinen}(2016)}]{Laine:2016hma}%
  \BibitemOpen
  \bibfield  {author} {\bibinfo {author} {\bibfnamefont {M.}~\bibnamefont
  {Laine}}\ and\ \bibinfo {author} {\bibfnamefont {A.}~\bibnamefont
  {Vuorinen}},\ }\href {https://doi.org/10.1007/978-3-319-31933-9} {\emph
  {\bibinfo {title} {{Basics of Thermal Field Theory}}}},\ Vol.\ \bibinfo
  {volume} {925}\ (\bibinfo  {publisher} {Springer},\ \bibinfo {year} {2016})\
  \Eprint {https://arxiv.org/abs/1701.01554} {arXiv:1701.01554 [hep-ph]}
  \BibitemShut {NoStop}%
\bibitem [{\citenamefont {Fisher}(1978)}]{Fisher:1978pf}%
  \BibitemOpen
  \bibfield  {author} {\bibinfo {author} {\bibfnamefont {M.~E.}\ \bibnamefont
  {Fisher}},\ }\bibfield  {title} {\bibinfo {title} {{Yang-Lee Edge Singularity
  and phi**3 Field Theory}},\ }\href
  {https://doi.org/10.1103/PhysRevLett.40.1610} {\bibfield  {journal} {\bibinfo
   {journal} {Phys. Rev. Lett.}\ }\textbf {\bibinfo {volume} {40}},\ \bibinfo
  {pages} {1610} (\bibinfo {year} {1978})}\BibitemShut {NoStop}%
\bibitem [{\citenamefont {Stephanov}(2006)}]{Stephanov:2006dn}%
  \BibitemOpen
  \bibfield  {author} {\bibinfo {author} {\bibfnamefont {M.~A.}\ \bibnamefont
  {Stephanov}},\ }\bibfield  {title} {\bibinfo {title} {{QCD critical point and
  complex chemical potential singularities}},\ }\href
  {https://doi.org/10.1103/PhysRevD.73.094508} {\bibfield  {journal} {\bibinfo
  {journal} {Phys. Rev. D}\ }\textbf {\bibinfo {volume} {73}},\ \bibinfo
  {pages} {094508} (\bibinfo {year} {2006})},\ \Eprint
  {https://arxiv.org/abs/hep-lat/0603014} {arXiv:hep-lat/0603014} \BibitemShut
  {NoStop}%
\bibitem [{\citenamefont {Alm\'asi}\ \emph {et~al.}(2019)\citenamefont
  {Alm\'asi}, \citenamefont {Friman}, \citenamefont {Morita},\ and\
  \citenamefont {Redlich}}]{Almasi:2019bvl}%
  \BibitemOpen
  \bibfield  {author} {\bibinfo {author} {\bibfnamefont {G.~A.}\ \bibnamefont
  {Alm\'asi}}, \bibinfo {author} {\bibfnamefont {B.}~\bibnamefont {Friman}},
  \bibinfo {author} {\bibfnamefont {K.}~\bibnamefont {Morita}},\ and\ \bibinfo
  {author} {\bibfnamefont {K.}~\bibnamefont {Redlich}},\ }\bibfield  {title}
  {\bibinfo {title} {{Fourier coefficients of the net baryon number density and
  their scaling properties near a phase transition}},\ }\href
  {https://doi.org/10.1016/j.physletb.2019.04.023} {\bibfield  {journal}
  {\bibinfo  {journal} {Phys. Lett. B}\ }\textbf {\bibinfo {volume} {793}},\
  \bibinfo {pages} {19} (\bibinfo {year} {2019})},\ \Eprint
  {https://arxiv.org/abs/1902.05457} {arXiv:1902.05457 [hep-ph]} \BibitemShut
  {NoStop}%
\bibitem [{\citenamefont {Mukherjee}\ and\ \citenamefont
  {Skokov}(2021)}]{Mukherjee:2019eou}%
  \BibitemOpen
  \bibfield  {author} {\bibinfo {author} {\bibfnamefont {S.}~\bibnamefont
  {Mukherjee}}\ and\ \bibinfo {author} {\bibfnamefont {V.}~\bibnamefont
  {Skokov}},\ }\bibfield  {title} {\bibinfo {title} {{Universality driven
  analytic structure of the QCD crossover: radius of convergence in the baryon
  chemical potential}},\ }\href {https://doi.org/10.1103/PhysRevD.103.L071501}
  {\bibfield  {journal} {\bibinfo  {journal} {Phys. Rev. D}\ }\textbf {\bibinfo
  {volume} {103}},\ \bibinfo {pages} {L071501} (\bibinfo {year} {2021})},\
  \Eprint {https://arxiv.org/abs/1909.04639} {arXiv:1909.04639 [hep-ph]}
  \BibitemShut {NoStop}%
\bibitem [{\citenamefont {Connelly}\ \emph {et~al.}(2020)\citenamefont
  {Connelly}, \citenamefont {Johnson}, \citenamefont {Rennecke},\ and\
  \citenamefont {Skokov}}]{Connelly:2020gwa}%
  \BibitemOpen
  \bibfield  {author} {\bibinfo {author} {\bibfnamefont {A.}~\bibnamefont
  {Connelly}}, \bibinfo {author} {\bibfnamefont {G.}~\bibnamefont {Johnson}},
  \bibinfo {author} {\bibfnamefont {F.}~\bibnamefont {Rennecke}},\ and\
  \bibinfo {author} {\bibfnamefont {V.}~\bibnamefont {Skokov}},\ }\bibfield
  {title} {\bibinfo {title} {{Universal Location of the Yang-Lee Edge
  Singularity in $O(N)$ Theories}},\ }\href
  {https://doi.org/10.1103/PhysRevLett.125.191602} {\bibfield  {journal}
  {\bibinfo  {journal} {Phys. Rev. Lett.}\ }\textbf {\bibinfo {volume} {125}},\
  \bibinfo {pages} {191602} (\bibinfo {year} {2020})},\ \Eprint
  {https://arxiv.org/abs/2006.12541} {arXiv:2006.12541 [cond-mat.stat-mech]}
  \BibitemShut {NoStop}%
\bibitem [{\citenamefont {Wood}(1992)}]{dwood}%
  \BibitemOpen
  \bibfield  {author} {\bibinfo {author} {\bibfnamefont {D.}~\bibnamefont
  {Wood}},\ }\href {http://www.cs.kent.ac.uk/pubs/1992/110} {\emph {\bibinfo
  {title} {The Computation of Polylogarithms}}},\ \bibinfo {type} {Tech. Rep.}\
  \bibinfo {number} {15-92*}\ (\bibinfo  {institution} {University of Kent,
  Computing Laboratory},\ \bibinfo {address} {University of Kent, Canterbury,
  UK},\ \bibinfo {year} {1992})\BibitemShut {NoStop}%
\end{thebibliography}%

\end{document}